%% file: s190814bv.tex
\documentclass[twocolumn]{aastex63}

\usepackage{amsmath}
\usepackage{amssymb}
\usepackage{amsthm}
\usepackage{graphicx}
\usepackage[colorinlistoftodos]{todonotes}
\usepackage{natbib}

\usepackage{lineno}


\usepackage{subfigure}
\usepackage{soul}
\usepackage{verbatim}

\setwatermarkfontsize{0.75in}
\definecolor{darkgreen}{rgb}{0.05,0.3,0.05}

\begin{document}

\vspace*{-\headsep}\vspace*{\headheight}
{\footnotesize \hfill FERMILAB-PUB-20-219-AE}\\
\vspace*{-\headsep}\vspace*{\headheight}
{\footnotesize \hfill DES-2019-0501}

\title{Constraints on the Physical Properties of GW190814 through Simulations based on DECam Follow-up Observations by the Dark Energy Survey}

\correspondingauthor{Robert Morgan}
\email{robert.morgan@wisc.edu}

\author[0000-0002-7016-5471]{R.~Morgan}
\affil{Physics Department, University of Wisconsin-Madison, 1150 University Avenue Madison, WI  53706, USA}
\affil{Legacy Survey of Space and Time Corporation Data Science Fellowship Program, USA}

\author[0000-0001-6082-8529]{M.~Soares-Santos}
\affil{Brandeis University, Physics Department, 415 South Street, Waltham MA 02453}

\author[0000-0002-0609-3987]{J.~Annis}
\affil{Fermi National Accelerator Laboratory, P. O. Box 500, Batavia, IL 60510, USA}

\author[0000-0001-6718-2978]{K.~Herner}
\affil{Fermi National Accelerator Laboratory, P. O. Box 500, Batavia, IL 60510, USA}

\author[0000-0001-9578-6322]{A.~Garcia}
\affil{Brandeis University, Physics Department, 415 South Street, Waltham MA 02453}

\author[0000-0002-6011-0530]{A.~Palmese}
\affil{Fermi National Accelerator Laboratory, P. O. Box 500, Batavia, IL 60510, USA}
\affil{Kavli Institute for Cosmological Physics, University of Chicago, Chicago, IL 60637, USA}

\author[0000-0001-8251-933X]{A.~Drlica-Wagner}
\affil{Fermi National Accelerator Laboratory, P. O. Box 500, Batavia, IL 60510, USA}
\affil{Kavli Institute for Cosmological Physics, University of Chicago, Chicago, IL 60637, USA}
\affil{Department of Astronomy and Astrophysics, University of Chicago, Chicago, IL 60637, USA}

\author[0000-0003-3221-0419]{R.~Kessler}
\affil{Kavli Institute for Cosmological Physics, University of Chicago, Chicago, IL 60637, USA}
\affil{Department of Astronomy and Astrophysics, University of Chicago, Chicago, IL 60637, USA}

\author[0000-0002-9370-8360]{J.~Garc\'ia-Bellido}
\affil{Instituto de Fisica Teorica UAM/CSIC, Universidad Autonoma de Madrid, 28049 Madrid, Spain}

\author[0000-0002-6119-5353]{T.~G.~Bachmann}
\affil{Department of Astronomy and Astrophysics, University of Chicago, Chicago, IL 60637, USA}

% ...

\begin{abstract}
On 14 August 2019, the LIGO and Virgo Collaborations detected gravitational waves from a black hole and a 2.6 solar mass compact object, possibly the first neutron star -- black hole (NSBH) merger.
In search of an optical counterpart, the Dark Energy Survey (DES) obtained deep imaging of the entire 90 percent confidence level localization area with Blanco/DECam 0, 1, 2, 3, 6, and 16 nights after the merger. 
Objects with varying brightness were detected by the DES Pipeline and we systematically reduced the candidate counterparts through catalog matching, light curve properties, host-galaxy photometric redshifts, SOAR spectroscopic follow-up observations, and machine-learning-based photometric classification.
All candidates were rejected as counterparts to the merger. 
To quantify the sensitivity of our search, we applied our selection criteria to full light curve simulations of supernovae and kilonovae as they would appear in the DECam observations.
Since the source class of the merger was uncertain, we  utilized an agnostic, three-component kilonova model based on tidally-disrupted NS ejecta properties to quantify our detection efficiency of a counterpart if the merger included a NS.
We find that if a kilonova occurred during this merger, configurations where the ejected matter is greater than 0.07 solar masses, has lanthanide abundance less than $10^{-8.56}$, and has a velocity between $0.18c$ and $0.21c$ are disfavored at the $2\sigma$ level. 
Furthermore, we estimate that our background reduction methods are capable of associating gravitational wave signals with a detected electromagnetic counterpart at the $4\sigma$ level in $95\%$ of future follow-up observations.
\newline
\end{abstract}


\section{Introduction}

The field of multimessenger astrophysics has experienced dramatic growth in the past few years thanks to the development and increased sensitivities of instruments like the Advanced Laser Interferometer Gravitational-Wave Observatory \citep[LIGO;][]{advanced_ligo}, the Virgo Interferometer \citep{Acernese_2014}, IceCube \citep{icecube}, and ANTARES \citep{antares}.
As well, real-time alert streams of detections made by these instruments such as Astrophysical Multimessenger Observatory Network \citep[AMON;][]{amon, amon_icrc} and the Gamma-ray Coordination Network \citep[GCN;][]{GCN_network} have made it possible for the astronomical community to target the sources of gravitational-waves \citep[][among several others from the previous LIGO/Virgo observing runs]{Doctor_2019, a20,  gwtarget10,  gwtarget11, collaboration2020desgw} and  high-energy neutrinos \citep[][among several others]{stein2020highenergy, me:), neutrinos_panstarrs, txs, neutrinos_old, neutrinos_old_2} in search of an electromagnetic signal within hours of the first detection of astrophysical events.

The most notable multimessenger observation to date is the association of the gravitational wave signal of two coalescing neutron stars (GW170817) detected by LIGO and Virgo \citep{gw170817_ligo}, a short gamma ray burst \citep[sGRB; GRB 170817A][]{gw170817_and_grb, integral_detection} detected by the Fermi Gamma Ray Burst Monitor \citep{fermigrbm} and INTernational Gamma-ray Astrophysics Laboratory \citep[INTEGRAL][]{integral}, and the observation of a kilonova (KN) AT2017gfo \citep{gw170817_1, gw170817_2, gw170817_3, gw170817_mm, kndiscovery, gw170817_hsc} in the nearby galaxy NGC 4993 \citep{Palmese_2017, gw170817_7}.
While this single event captured the focus of the entire astronomical community, the breadth and number of scientific analyses stemming from it are perhaps more astounding.
Standard siren techniques enabled a direct measurement of the expansion rate of the Universe today (\citealt{170817_siren}; \citealt*{des_darksirens}), and in the future they will reach a few percent precision on the Hubble constant \citep[]{Chen2018}, and on the growth of structure parameter $f\sigma_8$ when galaxies' peculiar velocities are used \citep[]{palmese20}.
Measuring element abundances in the merger ejecta using spectroscopic instruments led to an understanding of the origin of heavy elements synthesized during the merger \citep[]{gw170817_4, Tanaka_2018, Drout1570}.
X-ray and radio observations characterized the geometry of the explosion to be best-described by a jet plus cocoon structure \citep{gw170817_5, gw170817_6}.
The gravitational waveforms tested and further evidenced the theory of General Relativity \citep[]{PhysRevLett.123.011102}, as verified by numerical relativity simulations \citep{num_rel_1, num_rel_2}.
\cite{gw170817_8}, \cite{Palmese_2017}, \cite{10.1093/mnras/stz891}, \cite{Lyman2018}, and several others explored the connection between binary neutron star (BNS) mergers and sGRBs.
All of these analyses, and many others not listed, were enabled by the association of the GW signal with its electromagnetic signal.
Needless to say, finding counterparts to gravitational waves from compact object mergers remains a primary goal of the multimessenger-focused astronomical community.

On 2019 August 14, two years later, LIGO and Virgo reported a candidate gravitational wave (GW190814) from another interesting compact object coalescence, and classified the source as a 23.2~$M_\sun$ black hole merging with a 2.6~$M_\sun$ compact object \citep{Abbott_2020_0814, GCN_324, GCN_333}. 
With the nonzero probability that this merger contained a neutron star, GW170814 again drew the interest of the astronomical community as potentially the first detected neutron star -- black hole (NSBH) merger.
The presence of a BH in the merger could significantly alter the electromagnetic signal compared to the previously discovered BNS event.
The electromagnetic signal is emitted by the ejection of NS matter during the coalescence, the characteristics of which strongly depend on the dynamics of the merger.
By analytically examining the tidal disruption of NSs by BHs,~\cite{lattimer_tidal_1976} found that only certain configurations of NSBH systems,
predominantly those with $M_{BH} \lessapprox 9$M$_{\sun}$, would produce ejections, and if they did, the resulting decompressing neutron star material would be rich in $r$-process nucleo-synthesis elements \citep[]{lattimer_decompression_1977, Capano2020}. 
The radioactive decay of these elements is expected to produce an optical counterpart usually referred to as ``kilonova'' or ``macronova'' \citep[]{metzger_electromagnetic_2010, barnes_effect_2013}, potentially similar for both NS and NSBH mergers \citep[]{Kawaguchi_2016, Fern_ndez_2017}. 
Increasingly sophisticated numerical simulations focused attention on the tidal structures of the ejected material \citep[]{rosswog_mass_1998, grossman_long-term_2014, 10.3389/fspas.2020.00046} and interesting physical mechanisms occurring in the event, such as neutrino-driven winds decompressing neutron star material \citep[]{perego_neutrino-driven_2014}.
In these simulations it was clear that not every merger produces a thick disk of material, and not every disk of material launches a jet that produces a GRB.
EM observations of events such as GW190814 can probe the dynamics of NSBH systems whether or not they identify a clear EM counterpart.

We, the Dark Energy Survey Gravitational Wave Search and Discovery Team (DESGW), targeted the localization area of GW190814 in search of an optical counterpart. 
We used the 4m Victor M. Blanco optical telescope and Dark Energy Camera imager \citep{brenna} from the Cerro-Tololo Inter-American Observatory in Chile.
Our observations tiled the areas of highest probability on each of the first four nights following the merger to search for rapidly-evolving transients, and then again six and sixteen nights following the merger to develop light curves of all objects in the field.
Each night, our DECam observations covered the entire 90\% localization area, and reached comparable depth or deeper than all other follow-up teams issuing GCN circulars documenting their observations. 
The localization area falls entirely within the Dark Energy Survey (DES) footprint, a 5,000 sq deg region of the southern sky which has been observed with DECam over the course of 6 years \citep[]{des_finale}.
We searched for candidate EM counterparts by comparing the images collected during the real-time observations to archival DES data.
After each night of observations, we published GCN circulars containing lists of potential GW190814 counterparts to alert other telescopes of their presence \citep[]{GCN_1, GCN_2, GCN_3, GCN_4, GCN_5, GCN_6, GCN_7, GCN_8, GCN_9, GCN_10, GCN_11, GCN_12, GCN_13, GCN_14, GCN_15, GCN_16}.
The overlap with the DES footprint also enabled a statistical standard siren measurement using this event and the DES galaxies \citep{palmese2020statistical}.

We developed detailed simulations of various types of KNe and supernovae (SNe, the largest expected contaminant) as they would appear in our observations.
We utilized the simulations to tune our KN selection criteria and to analyze the numbers and properties of the objects that made it to our final candidate sample.
We also developed a machine learning (ML) classifier to distinguish between KN and SN light curves, and analyzed its performance on our simulations and candidates in parallel.
From this simulation-based sensitivity analysis of the real-time follow-up observations, we report the expected numbers of objects passing our selection criteria, the detection efficiency of different KN models in our follow-up observations, the mean light curves of SNe and KNe in our final candidate sample, constraints on the properties of the merger ejecta, and a statistical forecast of our KN discovery potential going forward.

To date, multiple groups have performed and documented optical follow-up observations of GW190814.
The Global Relay of Observatories Watching Transients Happen (GROWTH) utilized the public DECam images discussed in this work \citep[henceforth G20]{a20}, \cite{v20} (henceforth V20) collected independent follow-up observations with the MegaCam instrument  \citep{MegaCam} on the Canada-France-Hawaii Telescope, the ElectromagNetic counterparts of GRAvitational wave sources at the VEry Large Telescope (ENGRAVE) Collaboration performed their follow-up observations using the Very Large Telescope \citep[henceforth E20]{e20}, \cite{m20} (henceforth M20) carried out follow-up observations with a network of telescopes, and \cite{w20} (henceforth W20) utilized the DDOTI wide-field imager.
All analyses, in similar fashion to the work presented here, systematically reduce the set of optical counterpart candidates, conclude a non-detection of an EM counterpart to GW190814, and proceed to place constraints on optical and dynamical properties of the candidate NSBH merger.
We compare the methodology and results from these works to ours in Section~\ref{sec:discussion}.
Briefly, analysis leverages simulated light curves of KNe and SNe to more accurately estimate the detection efficiency of a potential counterpart and to also estimate the expected background under the candidate selection methodology applied during the real-time follow-up observations, while employing the deep imaging capabilities and wide field of view of Blanco/DECam.

This paper is organized as follows.
In Section~\ref{sec:lvc_obs}, we summarize the properties of the GW signal detected by LIGO and Virgo.
In Section~\ref{sec:decam_obs} we summarize the DECam observations, observing strategy, and image processing.
Section~\ref{sec:candidate_select} presents the selection criteria and results of the real-time follow-up observations.
From these selection criteria, we also present a full simulation-based sensitivity analysis of the DECam observations of GW190814, and characterize the types of objects expected in our final candidate sample in Section \ref{sec:sensitivity}.
In Section~\ref{sec:discussion}, we utilize the sensitivity analysis results to inform a discussion of the dynamics of the merger and to discuss efficient search strategies for future events.
In Section~\ref{sec:discussion}, we also compare our results to those of other analyses.
We conclude in Section~\ref{sec:conclusion}.
Appendix \ref{app:ml} provides a summary of the ML photometric classification methods utilized in this analysis.
For all cosmological calculations, we adopt a Flat $\Lambda$~CDM cosmology with $H_0$ = 70.0~km/sec/Mpc and $\Omega_m = 0.30$.

\begin{deluxetable*}{cccccccc}
\tablehead{
\colhead{Night} & \colhead{Epoch} & \colhead{Filter} & \colhead{Airmass} & \colhead{PSF FWHM} & \colhead{Sky} & \colhead{Cloud} & \colhead{$5\sigma$ Depth} \\ 
& & & & (arcsec) & ($\Delta$mag) & ($\Delta$mag) & (mag) 
}
\startdata
20190814&    00d 09:22:04&    i&    1.04&    1.34&    2.83&    0.32&    20.32\\ 
        &                &    z&    1.04&    1.34&    2.02&    0.44&    20.13\\ 
20190815&    01d 08:20:28&    i&    1.05&    1.32&    2.51&    0.24&    21.28\\ 
        &                &    z&    1.02&    1.27&    1.51&    0.26&    21.21\\ 
20190816&    02d 07:41:53&    i&    1.09&    1.93&    1.22&    0.04&    21.65\\ 
        &                &    z&    1.09&    1.83&    0.55&    0.06&    21.67\\ 
20190817&    03d 08:43:18&    i&    1.04&    1.13&    1.31&    0.01&    21.99\\ 
        &                &    z&    1.04&    1.10&    0.69&    0.03&    21.96\\ 
20190820&    06d 06:53:41&    i&    1.13&    1.06&    0.60&    0.09&    22.96\\ 
        &                &    z&    1.12&    1.04&    0.25&    0.11&    22.71\\ 
20190830&    16d 07:34:26&    i&    1.10&    0.94&   -0.42&    0.15&    23.76\\ 
\enddata
\caption{Summary of observing conditions during the DECam targeting of GW190814. The ``Sky'' and ``Cloud'' columns refer to the effect on the limiting magnitude of the observations resulting from background sky brightness and extinction due to cloud cover, respectively. 
\label{tab:obs}
}
\end{deluxetable*}

\begin{figure}[p]
\centering
\includegraphics[width=1.0\linewidth, trim={1cm 0.5cm 1cm 0.5cm}, clip]{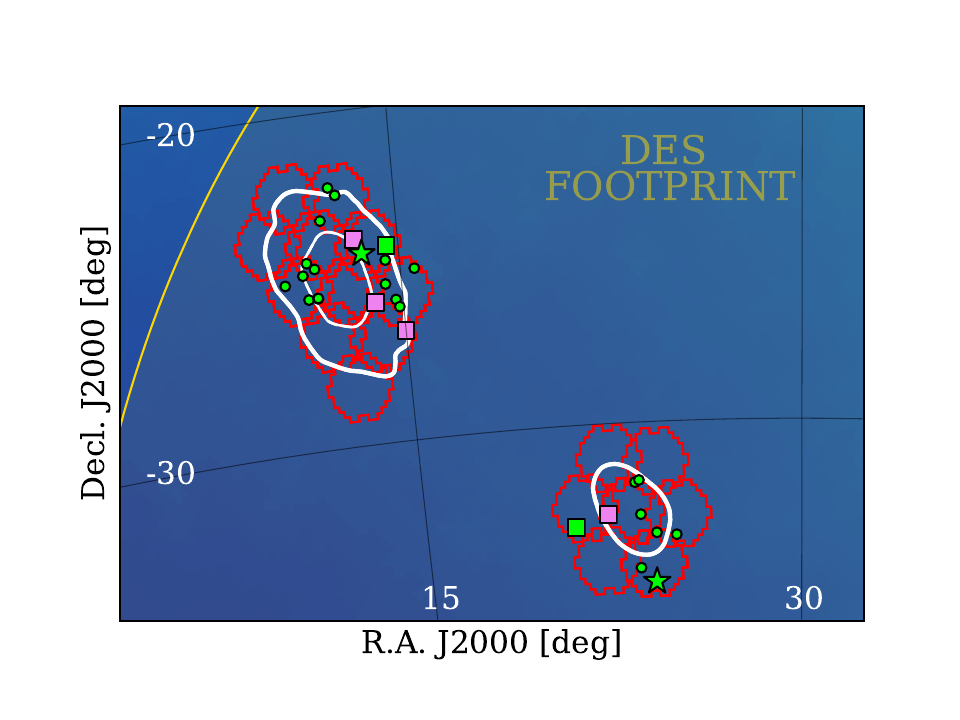}
\caption{Summary of exposures taken and candidates identified by the DESGW pipeline. 
DECam pointings are shown as orange hexes and represent the area covered on nights 2-5. 
Additional images were taken using a different tiling in order to eliminate chip gaps, those hexes are not shown for simplicity.
The white contours are the LVC $90\%$ (bold) and $50\%$ probability region. Finally, the gold line represents the boundary of the DES footprint.
Stars represent candidates that pass all selection criteria prior to final ML classification and have not been targetted with spectroscopic instruments.
Circles show candidates reported via GCN circulars that were ruled out in this analysis.
Squares denote candidates that were spectroscopically confirmed as SNe.
Violet coloring indicates a candidate was first reported by a group other than DESGW, while green coloring is used for DESGW candidates.
}
\label{fig:hexes}
\end{figure}

\begin{figure}
\centering
\includegraphics[width=1.0\linewidth, trim={0.4cm 0.6cm 0.4cm 0.1cm}, clip]{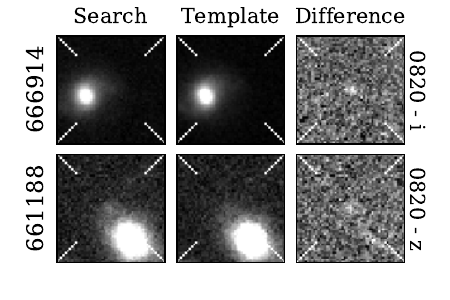}
\caption{Images of objects passing all selection criteria before machine-learning classification. For each object, the set of images for the night with the least noisy difference image is displayed. All images are centered on the detected transient. The DESGW ID of the object is listed on the left axis label, while the ``MMDD'' date and filter used are shown as the right axis label. Each image has dimensions $13.4\arcsec \times 13.4\arcsec$. \label{fig:stamps} }
\end{figure}

\section{LIGO/Virgo Observations}
\label{sec:lvc_obs}

On 2019 August 14, The LIGO/Virgo Consortium (LVC) reported the observation of gravitational radiation at high statistical significance \citep[]{Abbott_2020_0814}.
The event, named GW190814,  occurred when all three detectors (LIGO Hanford Observatory,  LIGO Livingston Observatory, and Virgo Observatory) were operating normally, which enabled both high precision localization of the source and more precise waveform parameter estimation.
The false alarm probability was calculated at $2.0\times 10^{-33}$~Hz, once per $10^{15}$~Hubble times, suggesting a very high signal-to-noise ratio event \citep[]{GCN_333}.  
The source of the GW signal was localized to a $38$ $(7)$~sq.~degree area at the $90\%$ ($50\%$) confidence level in the southern hemisphere on the night of the merger.
The localization area was split into two distinct regions, shown in Figure \ref{fig:hexes}, as a result of polarization and timing information from the three-detector detection.
Preliminary parameter estimation using the {\tt bayestar} pipeline classified the event as falling into the ``Mass-Gap'', meaning the detected GW was consistent with at least one of the objects having mass having mass between 3~$M_\sun$ and~5~$M_\sun$.
The small localization area and the presence of a low-mass compact object, potentially a massive neutron star, made this event interesting from the perspective of electromagnetic follow-up \citep[]{massgap}.
The following day, the LVC \texttt{LALInference} pipeline localized the source to $23 (5)$~sq.~degrees at the $90\%$ ($50\%$) confidence level, refined the classification to a NSBH merger, and estimated the luminosity distance of the event to be $267\pm52$~Mpc.
DECam follow-up observations proceeded based on this information, but in June 2020, the LVC released its final parameter estimation values for GW190814: the luminosity distance was revised to $239\pm43$~Mpc; the 90\% localization area was reduced to an $18.5$~sq.~degree section of the original 90\% localization area; and the masses of the objects involved in the merger were refined to 23.2~M$_\sun$ and 2.6~M$_\sun$ \citep[]{Abbott_2020_0814}.

\section{DECam Observations}
\label{sec:decam_obs}

In search of an EM counterpart to GW190814, we triggered Target of Opportunity (ToO) observations with the 4m Victor M. Blanco Telescope located at Cerro Tololo Inter-American Observatory in Chile.
The Blanco was equipped with with DECam, a 570-mega pixel optical imager \citep{brenna}.
Together, the Blanco and DECam reach a $5\sigma$ limiting $r$-band magnitude of $\sim23.5 \textrm{ mag}$ in a 90 second exposure in a 3 sq. deg field of view (FoV) \citep{des_dr1}.
The combination of deep imaging and a wide FoV make Blanco/DECam the ideal southern hemisphere instrument for efficiently detecting explosive optical transients localized to tens of square degrees.

Our follow-up efforts for GW190814 utilized the resources of the DES, which is a wide-field optical survey that covered a $\sim5,000$ sq. degree region (referred to henceforth as the DES footprint) of the southern sky from 2013 to 2019 using Blanco/DECam.
DES imaging of the DES footprint is expected to reach a $10\sigma$ co-added depth for point sources of $grizY$ = 24.7, 24.5, 23.8, 23.1, 21.9~mag.
The LVC 90\% containment region for GW190814 is entirely within the DES footprint, enabling the use of high-quality DES images during difference imaging.

\subsection{Observing strategy}
\label{sec:decam_obs_strategy}

We performed ToO follow-up observations of GW190814 0, 1, 2, 3, 6, and 16 nights following the LVC alert.
The early nights were chosen to look for rapidly evolving transients immediately following the merger, and the observations 16 nights after the merger were used to exclude persisting supernovae.
The observing conditions for each night are displayed in Table \ref{tab:obs}.

The moon was full on the first night of the observations, so we opted to use the redder $i$ and $z$ bands to minimize the effect of moon brightness on our imaging depth.
On the night of the merger, we tiled 99\% percent of the $38$ sq deg localization region using 60 second exposures in $i$ and 90 second exposures in $z$. The $z$ exposures were offset by half the width of a DECam CCD to fill in chip gaps.
We tiled the area a second time in $i$ to identify moving objects.
On the following observing nights, since the LVC had published a smaller localization region, we lengthened our exposures to 100 seconds in $i$ and 200 seconds in $z$.
Throughout the real-time observations, we coadded images that shared the same night and filter to increase the search depth.
The $i$-band DECam pointings are shown atop the LVC localization probability contours in Figure \ref{fig:hexes}. 
All DECam images were immediately made public and available for download from the National Science Foundation's NOIRLab.




\begin{deluxetable*}{cccccc}
\tablehead{
\colhead{DESGW ID} &  \colhead{TNS Name$^1$} & \colhead{GCN / ID} & \colhead{R.A. J2000 (deg)} & \colhead{Decl. J2000 (deg)} & \colhead{Outcome}} 
\startdata
666914 & 2019aaak & ae & 24.102867 & -34.766918 & ML Prob. SN = 0.92 \\
661188 & 2019aabz & af & 13.631072 & -24.286258 & ML Prob. SN = 0.86 \\
\hline
624921 & 2019nqq & 25373 / c & 20.95506 & -33.034762 & SOAR$^5$ SN-Ic, ML Prob SN = 0.99 \\
627288 & 2019obc & 25438 / q & 14.566764 & -24.139771 & GTC$^4$ SN-Ia, ML Prob. SN = 0.97 \\
628966 & 2019npv & GROWTH$^2$ & 13.384642 & -23.832904 & GMT$^6$ SN-Ibc, ML Prob SN = 0.97 \\
614750 & 2019nqc & GROWTH$^2$ & 22.265251 & -32.705166 & SALT$^3$ SN-II, ML Prob SN = 0.99 \\
661833 & 2019ntr & GROWTH$^2$ & 15.007796 & -26.714266 & SOAR$^5$ SN-Ia, ML Prob SN = 0.96 \\
626761 & 2019npw & GROWTH$^2$ & 13.968327 & -25.783283 & SOAR$^5$ SN-IIP, ML Prob. SN = 0.92 \\
\hline
614812 & 2019nmd & 25336 / a & 12.870848 & -22.471377 & Removed by Criterion 1 \\
614830 & 2019nme & 25336 / b & 12.635660 & -22.226027 & Removed by Criterion 1 \\
624609 & 2019nqr & 25373 / d & 23.573539 & -32.741781 & SOAR$^5$ SN-IIb, Removed by Criterion 4 \\
626209 & 2019nqs & 25373 / e & 23.396516 & -31.780134 & Removed by Criterion 6 \\
631484 & 2019nte & 25398 / f & 23.557358 & -31.7217 & Removed by Criterion 1 \\
635380 & 2019nxd & 25486 / i & 10.685824 & -24.955649 & Removed by Criterion 6 \\
625030 & 2019nxe & 25425 / j & 11.570058 & -24.372554 & Removed by Criterion 5 \\
625673 & 2019nys & 25486 / k & 14.487096 & -24.566822 & Removed by Criterion 6 \\
625633 & 2019nzr & 25425 / m & 11.839208 & -24.576827 & Removed by Criterion 4 \\
663323 & 2019oab & 25425 / o & 14.747491 & -25.770182 & Removed by Criterion 4 \\
624252 & 2019odc & 25486 / r & 11.507039 & -25.459150 & Removed by Criterion 5 \\ 
659801 & 2019okr & 25486 / t & 11.848733 & -25.458549 & Removed by Criterion 3 \\ 
625985 & 2019oks & 25486 / u & 15.534661 & -24.906027 & Removed by Criterion 5 \\
627394 & 2019omt & 25486 / v & 14.861426 & -25.994801 & GTC$^4$ SN-IIL, Removed by Criterion 6 \\
627577 & 2019omu & 25486 / w & 23.495376 & -34.338893 & Removed by Criterion 6 \\ 
627249 & 2019omv & 25486 / x & 24.978384 & -33.383719 & Removed by Criterion 6 \\ 
635566 & 2019omw & 25486 / y & 12.234396 & -23.170137 & Removed by Criterion 5 \\ 
625839 & 2019omx & 25486 / z & 24.18436 & -33.302678 & Removed by Criterion 5 \\ 
626718 & 2019onj & 25486 / ab & 11.858357 & -25.448647 & Removed by Criterion 3 \\ 
627832 & 2019opp & 25486 / ac & 14.409390 & -25.279166 & Removed by Criterion 6 \\ 
635044 & 2019aaah & ad & 11.382157 & -24.729753 & Removed by Criterion 6 \\
\enddata
\caption{Candidates found by the DESGW team during the real-time DECam observations of GW190814. 
Candidates found in real-time are listed in the bottom portion of the table, while candidates that pass the selection criteria developed in this work are listed in the top  portion. 
The middle section lists candidates that passed all criteria but were spectroscopically classified as SNe. The ``ML Prob. SN'' metric gives the probability that the object is a SN from the \texttt{PSNID + RFC} approach described in Section \ref{sec:candidate_select_ml} and Appendix \ref{app:ml}.
$^1$~The Transient Name Server, \url{https://wis-tns.weizmann.ac.il}. $^2$~Candidate first reported by the GROWTH Collaboration. $^3$~The Southern African Large Telescope. $^4$~The Gran Telescopio Canarias. $^5$~The Southern Astrophysical Research telescope. $^6$~The Giant Magellan Telescope. \label{tab:candidates}
}
\end{deluxetable*}

\subsection{Image Processing}
\label{sec:decam_obs_images}

The DECam images were processed by the DES Difference Imaging Pipeline, an updated version of the DES Supernova Program's Pipeline described in \citet{diffimg}, using coadded DES wide-field survey images as templates. 
The updated pipeline is described in detail in \citet{details} and has been used in a variety of multimessenger applications \citep{marcelle_15, gw170817_1, Doctor_2019, me:)}.
We describe our pipeline briefly below.

We apply standard image correction and  astrometric calibration to our DECam images \citep{des_desdm}, and subtract them from existing template images.
We utilize the GAIA-DR2 catalog \citep{2016A&A...595A...1G} to perform astrometric calibration to reach astrometric uncertainties smaller than $0.03$\arcsec.
After image subtraction, we use \texttt{SExtractor} \citep{sextractor} to locate sources in the difference images, which correspond to objects with varying brightness between the times of the template observations and the recent images.
We then obtain forced photometry at the locations of detected sources in the difference images using point-spread-function (PSF) fitting in which previous or future epochs have no detections.
Lastly, we apply a ML-based image artifact identification tool, {\tt autoscan} \citep{autoscan}, to the difference images to assign each detection a probability of being real as opposed to an artifact created by astrometric misalignment, hot/dead pixels, unidentified cosmic rays, etc.
All candidates found by the DESGW pipeline are listed in Table \ref{tab:candidates}.


\subsection{Host Galaxy Matching}
\label{sec:decam_obs_hostmatch}

We match each candidate to a host galaxy from the DES Y3 galaxy catalog.
After removing contaminants (subtraction artifacts, variable stars, moving objects, etc.) from our sample using criteria 1-5 described in Section \ref{sec:candidate_select}, every candidate is able to be matched to a host in the DES Y3 galaxy catalog.
Properties and redshifts of the hosts are reported in Table \ref{tab:hosts}. 
Photometric redshifts have been computed using Directional Neighborhood Fitting (DNF; \citealt{dnf}), while the galaxy properties have been computed using the method described in \citet{2019arXiv190308813P}. 
The DNF method is known to be inaccurate at the redshifts relevant in this analysis due to the characteristics of the galaxy sample upon which the algorithm was trained.
The inaccuracy manifests in our analysis as underestimated host galaxy photometric redshift uncertainty.
We therefore add a minimum uncertainty of 0.02 for galaxies with host galaxy photometric redshift less than 0.1 following the prescription of \citet*{des_darksirens}. 
The galaxies have been ranked from highest to lowest probability per unit volume based on their angular position and redshift as prescribed in \cite{2016ApJS..226...10S}, assuming a flat $\Lambda$CDM cosmology with $H_0=70$ km~s$^{-1}$ Mpc$^{-1}$ and $\Omega_m=0.3$.

\section{Candidate Selection}
\label{sec:candidate_select}

After the completion of our image processing pipeline, we found 33571 candidates.
The data sample includes astrophysical objects with varying brightness such as SNe, Active Galactic Nuclei (AGN), and other less-common explosive optical transients \citep[]{Cowperthwaite_2015}, moving objects such as minor planets and asteroids, foreground variable stars in the Milky Way, and image artifacts from poor image subtractions and insufficient masking of bright objects.
In the real-time analysis, we developed several selection criteria to look for the likely EM counterpart of the GW detection.
These selection criteria narrowed our sample to a size reasonable for spectroscopic, X-ray, and radio observing teams to follow up.
We detail those selection criteria here and evaluate their effectiveness at recovering KNe and rejecting background objects in the following section.

There are 9 selection requirements (criteria) in four levels: 
(1) subtraction quality requirements to reject image artifacts and moving objects, (2) catalog matching to rule out existing objects such as AGN and variable stars, (3) KN-specific requirements to rule out SNe, and (4) final candidate assessment using machine-learning (ML) based photometric classification. 
Each level progressively targets more specific properties of an expected EM counterpart.
The number of candidates remaining in our sample after each criterion are displayed in Table \ref{tab:cuts}.
The remainder of this section elaborates on the implementation and motivation for each selection criterion applied to the data.

\subsection{Level 1 Selection Criteria}
\label{sec:candidate_select_1}

The following selection criteria assure satisfactory detection and image-subtraction quality in all remaining candidates.
We introduce two definitions to expedite discussion. 
A Type-2 detection is a {\tt SExtractor} detection in a single filter that does not contain any image processing errors.
These errors include an inability to measure a fitted flux, the R.A. or Decl. of an object not being on a CCD, masking of bright objects overlapping the transient object, the inability to fit the PSF of the object, the inability to make a stamp in the difference image, a large number of pixels with negative flux values, and a $5\sigma$ difference between psf flux fitting and aperture flux fitting.
A Type-1 detection is a Type-2 detection that has also been given an {\tt autoscan} score of 0.7 or larger.

{\it Criterion 1.} We require candidates to have at least one Type-1 detection. 
This criterion ensures a high-purity sample of real objects with little contamination from image processing artifacts.

{\it Criterion 2.} We require a second detection in the light curve of Type-2 or Type-1, and we require this secondary detection to be on a different night from the detection in Criterion 1.
By ensuring a second detection that is separated in time from the first detection, we remove all moving objects from our sample.
This temporal separation could in principle be shortened to $\sim1$~hour, but because we co-added our images from the same night and band, this time separation requirement is effectively a multi-night requirement.
In these observations, we find that fast-fading transients such as KNe have a high efficiency of 93\% for this multi-night requirement based on the simulations discussed in Section \ref{sec:sensitivity}. 
We also relax the required {\tt autoscan} score of the second detection since the first Type-1 detection from Criterion 1 has already yielded a high-purity sample.


After the level-1 quality criteria, we are left with 2192 candidates in our sample.
This sample is mostly composed of astrophysical objects with observed variable brightness as a result of the quality criteria. 
There is a large population of artifacts still present at this stage that passed the selection criteria, but these are removed by Criterion 5. 

\subsection{Level 2 Selection Criteria}
\label{sec:candidate_select_2}

\begin{deluxetable*}{cccccccc}
\tablehead{\colhead{DESGW ID} & \colhead{Host Gal. Name} & \colhead{Angular Sep.} & Physical Sep. & Redshift & \colhead{$\log M_*$} & \colhead{$\log \textrm{SFR}$} & \colhead{M$_i$} \\  &  & [arcsec] & [kpc] &  & $[\log M_{\sun}]$ & $[\log M\sun / \textrm{yr}]$ &}
\startdata
666914 & DES J013624.60-344557.72 & 3.345 & 6.24 & 0.10 $\pm$ 0.02$^\dagger$ & 9.90 & -0.0386 & -20.70 \\
661188 & DES J005431.17-241713.08 & 4.700 & 9.53 & 0.11 $\pm$ 0.02$^\dagger$ & 10.07 & 0.0438 & -20.94 \\
\enddata
\caption{Host galaxy properties of the two objects passing all selection criteria prior to final ML classification. \\ $^\dagger$ The minimum host-galaxy photometric redshift uncertainty value has been utilized.
\label{tab:hosts}}
\end{deluxetable*}

With the exception of artifacts, we expect the remaining sample to be dominated by three main contaminants at this stage: variable foreground stars, AGN, and bright galactic centers. The latter is a known problem in difference imaging, see \cite{diffimg} or \cite{Doctor_2017} for context.


{\it Criterion 3.} We require that each object is separated from known foreground objects. 
This requirement has two components: each object must be separated from objects in a high purity sample of well-measured stars in the DES Y3 Gold catalog by at least $0.5\arcsec$, and each object must be separated at least $8\arcmin$ from NGC288 and $3\arcmin$ from HD4398. 
The globular cluster NGC288 has a high density of bright stars and HD4398 itself is a very bright star, both of which led to large numbers of subtraction artifacts and variable star detections by our Search and Discovery Pipeline.

{\it Criterion 4.} We require that each object is at least $0.2\arcsec$ from objects in the DES Y3 Gold catalog that are not flagged as well-measured stars, which were addressed in Criterion 3.
This criterion aims to remove AGN and bright galactic centers.
Section \ref{sec:sensitivity_criteria} gives physical and empirical motivations for expecting KNe to be highly likely to satisfy this requirement.

{\it Criterion 5.} We visually inspect images of the 1872 remaining candidates.
We remove candidates that have an imaging artifact from a misaligned subtraction or from inadequate masking and we also remove all candidates that contained a point-like light source in the template image at the location of the detected transient.
In the application of this criterion in general, the seeing of the observations can limit the efficiency of real transients, since extremely poor seeing could potentially make a bright host galaxy center appear as a point source.
Our average seeing in these observations, shown by the PSF FWHM column in Table \ref{tab:obs}, is less than 1.3\arcsec on more than half of the nights.
We therefore expect this behavior to be rare in our data.

After the level-2 catalog criteria, we are left with 116 candidates in our sample.
We expect that at this stage our data are almost entirely constituted by real astrophysical transients.

\subsection{Level 3 Selection Criteria}
\label{sec:candidate_select_3}

The following selection criteria are designed to remove supernovae by assuring the distance of the candidates is consistent with the LVC distance posterior distribution, requiring the light curves of the candidates are fading, and triggering spectroscopic follow-up observations.

{\it Criterion 6.} We require each object to have a host galaxy photometric redshift consistent with the mean and standard deviation of the LVC distance posterior at the $3\sigma$ confidence level.
All objects were able to matched with a host-galaxy at this stage, so the criterion can be straightforwardly applied.
The criterion is satisfied when
\begin{align}
    \frac{|z_\textrm{LVC} - z_\textrm{DES}|}{\sqrt{\sigma_{z,\textrm{LVC}}^2 + \sigma_{z,\textrm{DES}}^2}} < 3,
    \label{eq:redshift_cut}
\end{align}
where $z_\textrm{LVC} = 0.06$ is the redshift of GW190814, $z_\textrm{DES}$ is the redshift of a candidate's host galaxy, $\sigma_{z,\textrm{LVC}} = 0.005$ is the uncertainty on the redshift of GW190814, and $\sigma_{z,\textrm{DES}}$ is the uncertainty on the redshift of a candidate's host galaxy.
To implement this criterion, we utilize the assumed cosmology in this analysis.
In the case of an available spectroscopic redshift of the host galaxy, we utilize the spectroscopic information instead.
Since supernovae could be detectable out to large redshifts in these observations, we seek to remove contaminants in galaxies too distant to be associated with the GW signal.

{\it Criterion 7.} If an object is detected on the final night of observations (16 nights post-merger) we require that it be fainter than 22.5 mag in at least one band.
If an object is not detected on the 16th night, it passes this criterion.
This criterion removes rising and flat light curves from our candidate list.

{\it Criterion 8.} We trigger spectroscopic follow-up observations from the Southern Astrophysical Research \citep[SOAR;][]{soar} telescope on as many of the 8 remaining candidates as possible.
We also incorporate real-time spectroscopic classifications from other instruments during the follow-up based on circulars posted to the GCN.
The spectroscopic instruments were triggered in real-time, as opposed to after the selection criteria had been refined in the offline analysis, so there is not perfect overlap between the targeted objects and the remaining candidates presented in this work.
All targeted candidates were spectroscopically confirmed as SNe.

\subsection{Final Candidate Assessment}
\label{sec:candidate_select_ml}

\begin{figure*}
\centering
\includegraphics[width=0.49\linewidth]{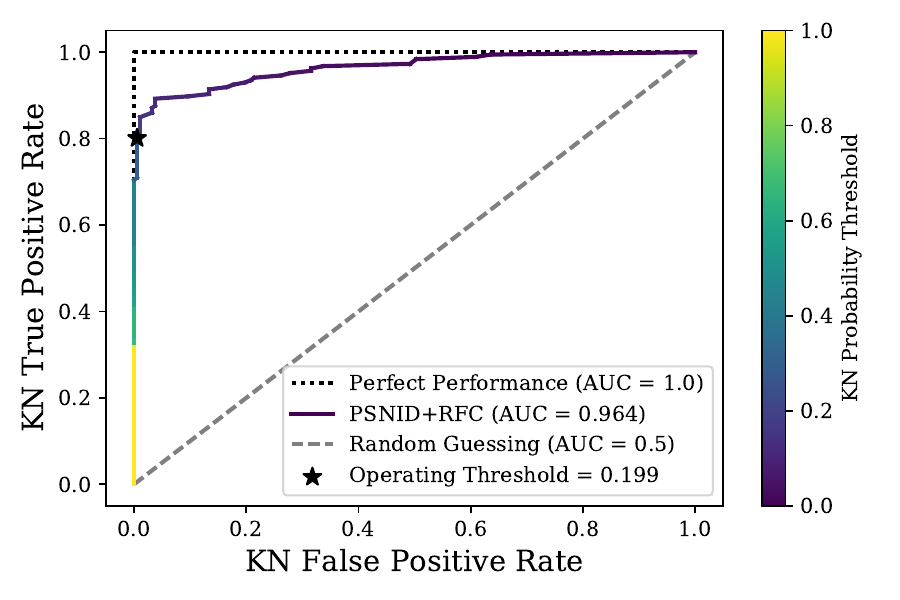}
\includegraphics[width=0.49\linewidth]{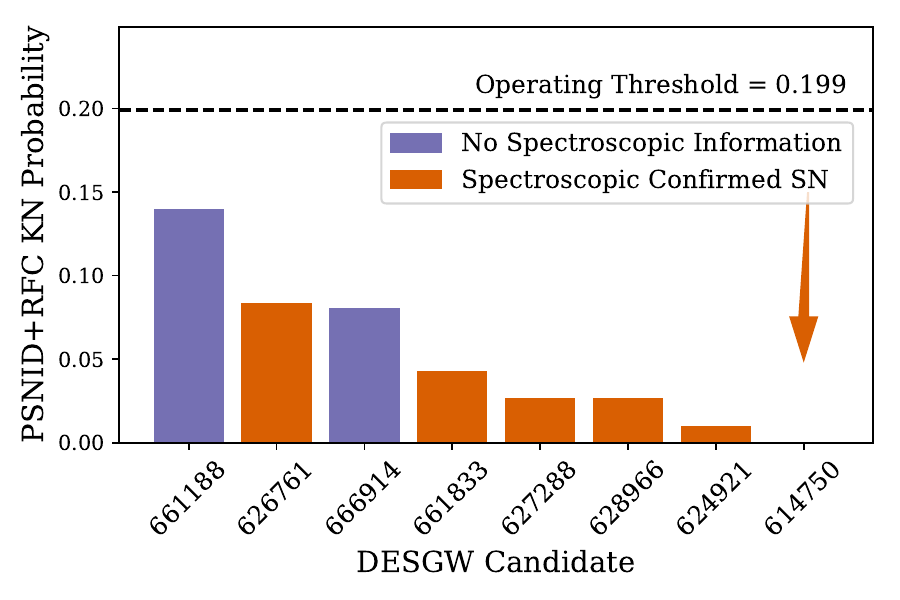}
\caption{Photometric classification of candidates using the {\tt PSNID + RFC} approach. \textit{Left:} Receiver Operating Characteristic Curves showing classification power of the {\tt PSNID + RFC} approach.
The threshold at which we chose to operate the classifier is denoted by a black star, the location of which shows the false positive rate and true positive rate of our ML approach. 
\textit{Right:} Calibrated probabilities of candidates passing Criterion 7. \label{fig:classification}}
\end{figure*}

After the previous eight criteria have been enforced, we have two remaining candidates as shown in Figure \ref{fig:stamps}. 
As described in Appendix \ref{app:ml}, we apply light-curve-based ML classification to determine the probability that any of these objects are potentially a KN.
Briefly, we fit a large set of simulated SNe (both SNe-Ia and SNe-CC) and KNe (from the \cite{Kasen2017} models) light curves that pass Criteria 1 through 7 
with a Bayesian SN template fitting tool {\tt PSNID} \citep{psnid}, select the template features and goodness of fit metrics with the largest difference in mean value for SN and KN samples, and build a random forest classifier \citep{randomforest} using those best-fit parameters as features.
This {\tt PSNID+RFC} approach shows a significant improvement in classification power when using the KN false positive rate and KN true positive rate as diagnostics.
A similar version of this method is described in \cite{me:)}.
Figure \ref{fig:classification} shows the performance of this machine learning approach and the resulting probabilities of each remaining candidate being a KN.
DESGW-666914 has a 0.92 probability of being a SN and DESGW-661188 has a 0.86 probability of being a SN from our \texttt{PSNID+RFC} approach, both of which are classified as SNe based on our choice of operating threshold. 
Six additional candidates that made it to this stage and were later spectroscopically typed as SNe were correctly classified as SNe by our \texttt{PSNID + RFC} approach.

Table \ref{tab:cuts} shows the number of candidates remaining after each criterion.
After all selection criteria have been applied and the remaining candidates have been photometrically classified, zero candidates remain.
We therefore use our data to set upper limits on KN properties given a non-detection and to inform future follow-up observations.

\section{Sensitivity Analysis}
\label{sec:sensitivity}

\begin{deluxetable*}{llccc}
\tablehead{
\colhead{No.} & \colhead{Description of Criterion} &
\colhead{Candidates} &  \colhead{Sim. SNe-Ia} & \colhead{Sim. SNe-CC}
}
\startdata
0 & DES Difference Imaging Pipeline & 33571 & 768.3  & 1191.1  \\ 
1 & Single Type-1 Detection & 2563 & 200.6  & 86.33 \\ 
2 & Two Type-2 Detections on Different Nights & 2192 & 118.8  & 48.29  \\ 
\hline
3 & Separated from Foreground Objects & 2021 & 117.8  & 47.9 \\ 
4 & $> 0.2\arcsec$ from DES Y3 GOLD Catalog Galaxy Centers & 1872 & 96.7  & 42.0  \\ 
5 & Visual Inspection of Stamps & 116 & 85.2  & 38.1  \\ 
\hline
6 & Redshift Consistent with LVC within 3 Standard Deviations & 9 & 4.7  & 6.5  \\ 
7 & Fainter than 22.5 mag on Night 16 & 8 & 2.6  & 4.8  \\ 
8 & Not Eliminated by Spectroscopic Observations & 2 & 0.8  & 1.4  \\ 
\hline
9 & Machine Learning Photometric Classification & 0 & $0.008$ & $0.014$ \\
\enddata
\caption{The selection criteria developed in this analysis and remaining objects after each criterion. The candidates column refers to objects found by the DES Difference Imaging Pipeline and the latter two columns show the expected number of SNe present in the candidate sample at each level computed as described in Section \ref{sec:sensitivity_sims}. The SNe simulations were realized 500 times so statistical uncertainty is negligible. The horizontal dividers reflect the ``levels`` of selection criteria described in the text. \label{tab:cuts}}
\end{deluxetable*}

To evaluate the selection criteria applied during the real-time observations, we model our search and selection methodology on simulated SNe and KNe using the SuperNova ANAlysis software suite \citep[\texttt{SNANA};][]{snana}.
The SNe and KNe models employed here are the same models used in the Photometric LSST Astronomical Time-series Classification Challenge \citep[PLAsTiCC;][]{plasticc}.
The SNe templates are derived from observations while the KNe templates are generated from theoretical models.
\texttt{SNANA} incorporates the cadence, the measured zeropoints, and noise level in the search and template images from our observations into the simulated fluxes and uncertainties to produce realistic light curves.
This simulation process enables the application of our real-time selection criteria to simulations and DECam candidates for a better understanding of what objects and how many of them would be expected to pass our selection criteria.
In the remainder of this section, we describe the {\tt SNANA} simulations for the GW190814 observations, detail the modeling of the selection criteria in the context of the simulations, and present the results of our sensitivity analysis: detection efficiencies for 329 different KN models, expected numbers of SNe to pass our selection criteria, the mean light curves of objects passing our selection criteria, upper limits on physical properties of potential EM counterparts to the GW190814 merger, and statistical forecasting of our discovery potential in follow-up observations of future events.

\subsection{Simulating the DECam Search}
\label{sec:sensitivity_sims}

\begin{figure*}
\centering
\includegraphics[width=0.82\linewidth, trim={0.2cm 0.5cm 0.2cm 1.5cm},clip ]{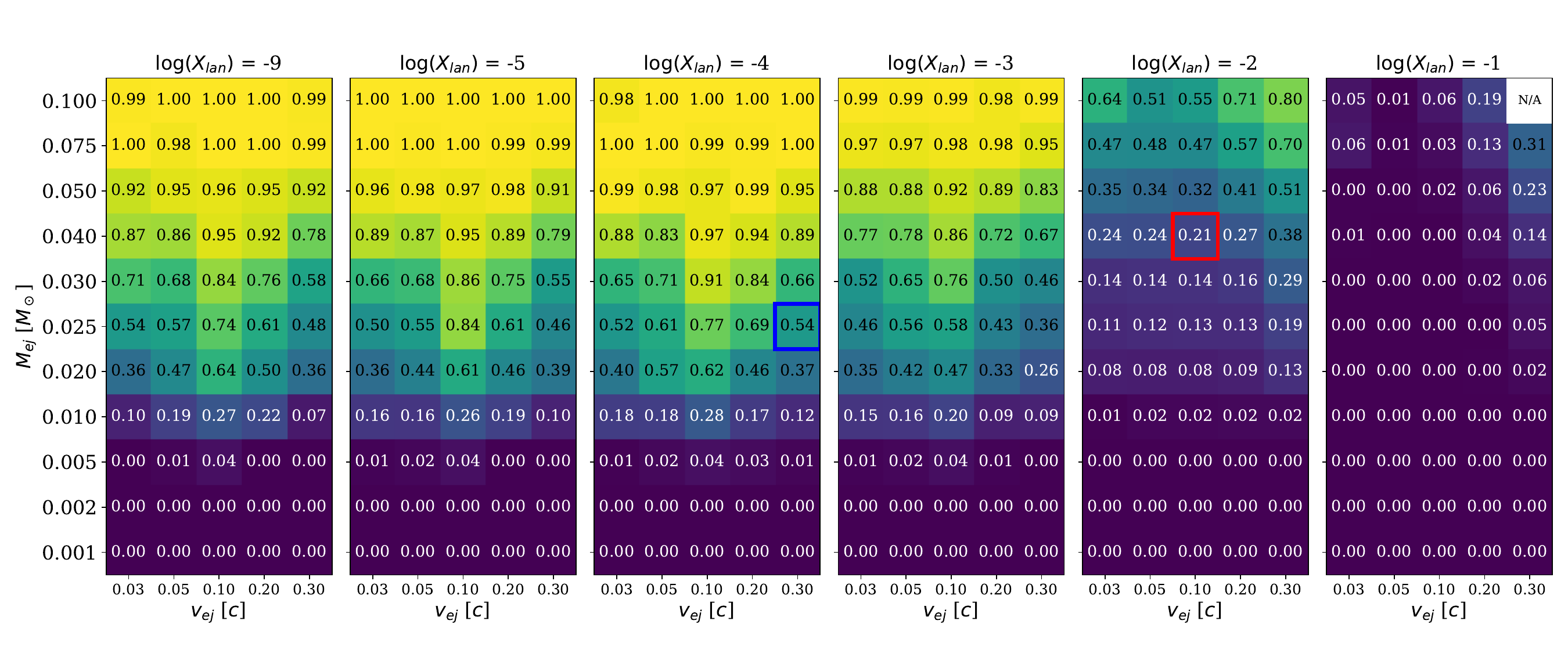}
\includegraphics[width=0.82\linewidth, trim={0.2cm 0.5cm 0.2cm 1.5cm},clip]{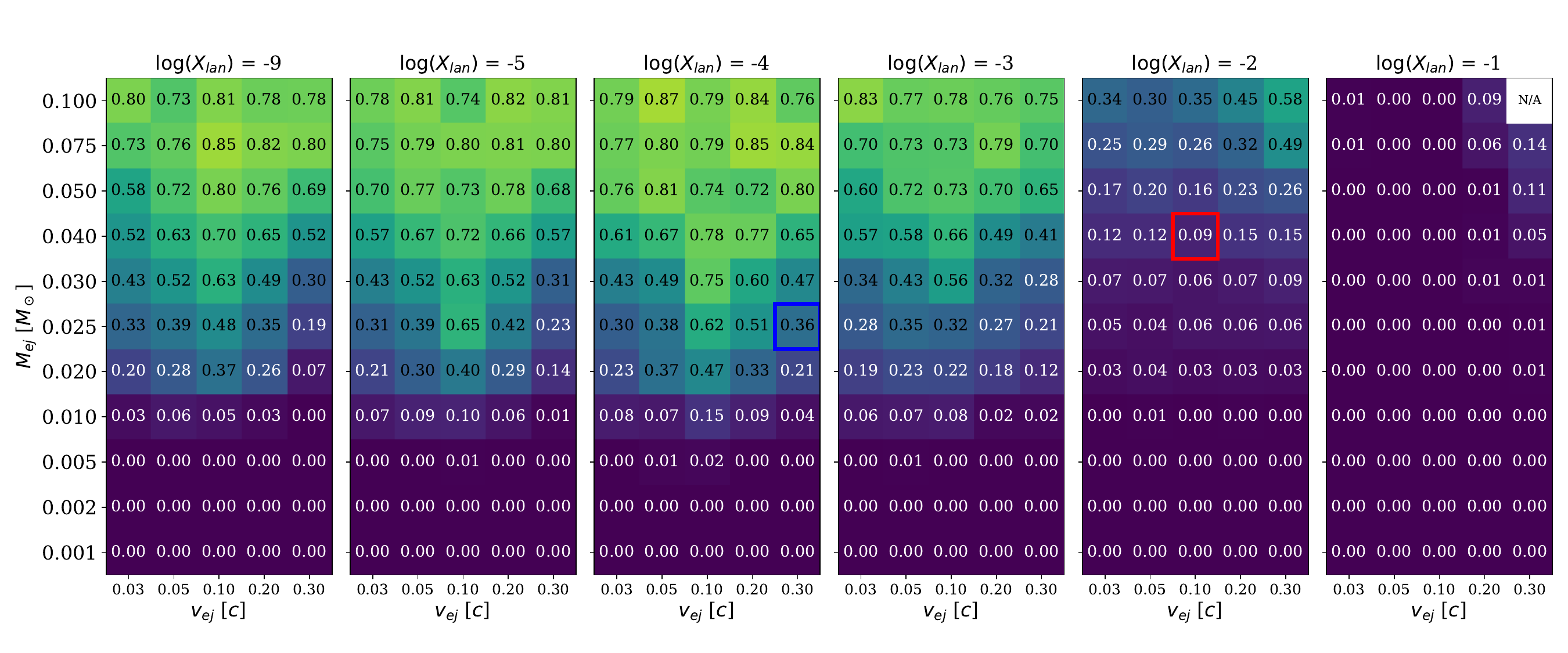}
\includegraphics[width=0.82\linewidth, trim={0.2cm 0.5cm 0.2cm 1.5cm},clip]{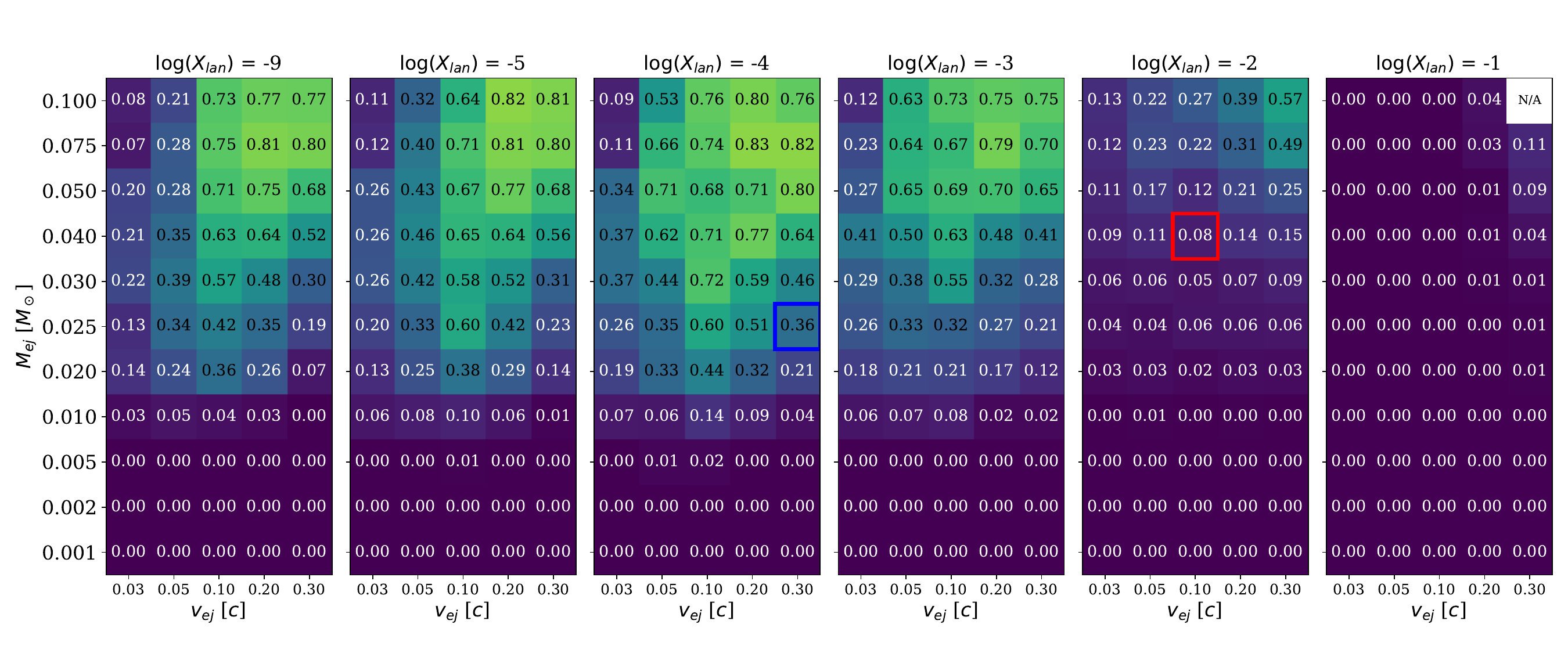}
\caption{KN efficiency parameterized by $M_{ej}$, $v_{ej}$, and $X_{lan}$. 
The blue and red boxes correspond to the blue and red components of the best-fit model for a GW170817-like KN at the distance of GW190814.
The top row shows the efficiencies of KN light curves in our observations after Criterion 1,  the middle row displays the efficiencies after all non-ML selection criteria have been applied, and the final row shows the efficiencies after the ML classification. 
All efficiencies are marginalized over the full LVC luminosity distance posterior distribution. 
The parameter combination marked ``N/A'' is  excluded because it is believed to be not well-motivated physically \citep[]{Kasen2017}.
}
\label{fig:KN_efficiency}
\end{figure*}

{\tt SNANA} enables the simulation of light curves of SNe, KNe, and other transients as they would be measured by DECam during observations.
This process uses a measured or theoretical time-evolving spectrum for the transient object and then accounts for cosmological redshift, Milky Way dust extinction, and the measured observing conditions of the DECam observations such as sky brightness, zeropoints, the point-spread-function of the imager, and CCD noise in the camera.
The corrected time-evolving spectra are then multiplied by the transmission of the DECam filters and light curves are sampled at epochs matching the cadence of the observations.

The KN models used in the simulations are from spectral energy distributions derived in \citet{Kasen2017} and parameterize the optical light from a KN by the mass ejected in the explosion, the abundance of lanthanide elements in the ejecta, and the velocity of the ejecta (hereafter $M_{ej}$, $X_{lan}$, and $v_{ej}$).
These models were chosen because they characterize the optical behavior based on physical properties of the NS ejecta, rather than having a dependence on the geometry or dynamics of the merger itself.
While other models for KNe and models specific to NSBH mergers exist
\citep[among multiple others]{nsbh1, nsbh2, nsbh3}, we find this simple, agnostic, three-component model based on observable properties of NS ejecta to apply well to GW190814.
In the simplicity of this approach, we make the assumption of either spherically symmetric emission or that the particular component being considered is directed towards Earth.
We utilize 329 total models, which discretize the parameter space in the ranges 0.001 $M_\sun$ $\leq M_{ej}  \leq $ 0.1 $M_\sun$, $0.03 c$ $\leq v_{ej} \leq$ $0.3 c$, and $1\times 10^{-9}$ $\leq X_{lan} \leq$ $1\times 10^{-2}$.
The simulated KNe are uniformly drawn from this population of models, hough the grid which discretizes the parameter space is non-uniform as shown in Figure \ref{fig:KN_efficiency}. 
This non-uniform grid is not believed to have an effect on the physical constraints insofar as the interpolation of model efficiencies between points in the grid is smooth and monotonic.
As a proxy for the observer-frame explosion time of the simulated KNe, we fix the time the KNe fluxes reach 1 percent of their peak flux to the time of the LVC GW alert and note that this approximation is justified by the rapid rise times of the KNe.
The simulated KNe are also distributed in redshift according to a polynomial fit of the LVC distance posterior and the cosmology used in this analysis.
The redshift distribution is constructed independent of spatial information on the sky.
This approximation is based on the small localization area of GW190814, however for future events with larger localization areas, the volume-rendered luminosity distance distribution should be utilized.

Because our selection criteria effectively remove all moving objects, known foreground variable stars, and AGN, the most likely remaining contaminants in our data are SNe.
We therefore use SN simulations to understand the types of SNe passing our selection criteria, as well as the number expected be present in our final candidate sample.
We simulate type-Ia SNe (SNe-Ia) using templates from \citep{guy2010salt2} and measured volumetric rates from \citep{Dilday_2008}.
We also simulate core-collapse SNe (SNe-CC) using templates from \citet{Kessler_2010} and volumetric rates from \citet{li2011_rates}.
The SNe-CC population includes type-Ib, type-Ic, type-Ibc, type-IIP, type-IIN, and type-IIL SNe, and we weight the different sub-types according to their measured volumetric rates.
Unlike the KN sample, we allow the SNe to have a random observer-frame explosion time that would make them bright enough to observe with DECam during our observing window.
This explosion time range is implemented by requiring the date of peak flux to be greater than 60 days prior to the LVC GW alert and less than 30 days after it, since the explosion time itself is not well-measured.

\subsection{Modeling Selection Criteria}
\label{sec:sensitivity_criteria}

{\tt SNANA} produces catalog-level photometric fluxes for transient objects by correcting model spectral energy distributions and multiplying them with the DECam filters, and this approach bypasses several image processing and catalog matching steps that we apply to the real DECam data.
We therefore take additional steps to impute information necessary for modeling the selection criteria in this analysis on the simulated light curves. 

In our real-time analysis, we applied selection requirements on the {\tt autoscan} score and {\tt SExtractor} detection flag.
Both of these programs run at the image level, so their information is not present in {\tt SNANA}-simulated light curves.
We adopt the empirical approach from \cite{Doctor_2017} to determine realistic values for {\tt autoscan} and {\tt SExtractor} quantities in the simulations.
This process involves inserting simulated point source objects of known brightness (hereafter ``fakes'') into the real DECam images, and applying our image processing pipeline to the images to record the {\tt autoscan} score and {\tt SExtractor} detection flag.
From the processed fake objects, we extract the probability mass functions (pmfs) for the {\tt autoscan} score and {\tt SExtractor} detection flag at discrete levels of signal-to-noise ratio ranging from 0.5 to 50.0.
Each filter is treated independently when extracting the pmfs.
In the process of generating simulations, based on the signal-to-noise ratio of each observation, values for the {\tt autoscan} score and {\tt SExtractor} detection flag are drawn from the corresponding empirically-derived pmf.
We also introduce a reduced correlation coefficient of 0.1 to the drawn {\tt autoscan} scores for observations of the same object, determined so that the simulations accurately reflect the fake data.

Level 2 of our selection criteria rules out known objects by matching to the DES Y3 Gold Catalog.
When matching to DES stars, globular clusters NGC288, and the star HD4398, we estimate the sky area masked by our criteria using a Monte-Carlo sampling of position space.
We find that a $0.5\arcsec$ radius around DES stars masks 0.11 percent of the sky area covered in our follow-up observations, and an $8\arcmin$ radius around NGC288 and a $3\arcmin$ radius around HD4398 each mask 0.01 percent.
In the simulations, we use these percentages of the sky masked by these selection criteria as the probability for a simulated object to be removed by the criterion.

We take a slightly different approach to modeling the criterion of removing known galactic centers from our sample, since these objects are not in the foreground of our observations.
Here we model the transient-galaxy separation empirically and impute that separation into the simulations.
We extract a probability distribution function of SN-host galaxy center separation in units of physical distance from the DES 3-year spectroscopic SNe sample \citep{DESSN}.
This sample is dominated by SN-Ia for cosmological analyses, which makes it more applicable to KN-host galaxy separation than a balanced SNe sample: the progenitors of SN-Ia are thought to be white dwarf stars in binary systems \citep[]{snprob1, snprops2, snprops3}, meaning to first order they would be similar in age and hence host separation to other binary systems of stellar remnants \citep[]{Bloom_2006, Prochaska_2006}.
We believe this assumption to be conservative, given that supernova (or sometimes called ``prenatal'') kicks during the evolution of binary massive star systems into BNS or NSBH systems are expected to cause an increase in the separation from the host-galaxy center \citep[]{kicks}.
We therefore apply the same transient-galaxy separation pdf to both the KNe and SNe simulations.
In the application of the selection criterion, we draw a separation from the pdf and remove the object if the separation is less than $0.2\arcsec$.
When testing this criterion on the DES 3-year SNe sample, we estimate 97\%\ of transients will be recovered while effectively removing all time-varying galactic centers.

Our real-time candidate reduction also relied on visual inspection of the images to remove artifacts and point-like light sources without a host galaxy.
We assume near perfect efficiency in the simulations with one exception stemming from the fact that this criterion has a dependence on the seeing of the observations.
A bright galaxy center in poor seeing conditions can hide real transients in the image or appear like a point source itself, resulting in it being removed from the sample.
For the simulations, if the imputed host separation is less than half of the seeing, we reject the simulated object.

The final pieces of additional information that were necessary to add to the {\tt SNANA} simulations were photometric redshifts and photometric redshift errors.
Here we take the $i$-band galaxy magnitudes of all galaxies in the DES Y3 Gold catalog also in the LVC 90\% containment region to empirically determine the $i$-band magnitude pdf in several redshift bins.
Using the true simulated redshift of our SNe and KNe, we select the corresponding host galaxy $i$-band magnitude pdf and draw a random value.
With a chosen $i$-band host magnitude, we determine the expected value of the photometric redshift error from the validation of the Gold catalog. 
We define a Gaussian distribution centered on the true simulated redshift with a standard deviation of the photometric redshift error.
We account for known underestimations of low redshift galaxies' photometric redshift uncertainty using the same treatment discussed in Section \ref{sec:decam_obs_hostmatch}. 
Thus, after drawing a photometric redshift from this distribution, each simulated transient will have a photometric redshift and photometric redshift error to match the candidates in our observations.

We model spectroscopic targeting and classification by implementing the ratio of the number of objects targeted by spectroscopic instruments to the number of candidates remaining at that stage in the follow-up as the probability of a SN being rejected.

\subsection{Sensitivity Results}
\label{sec:sensitivity_results}

\begin{figure*}[t]
\centering
\includegraphics[width=0.7\linewidth]{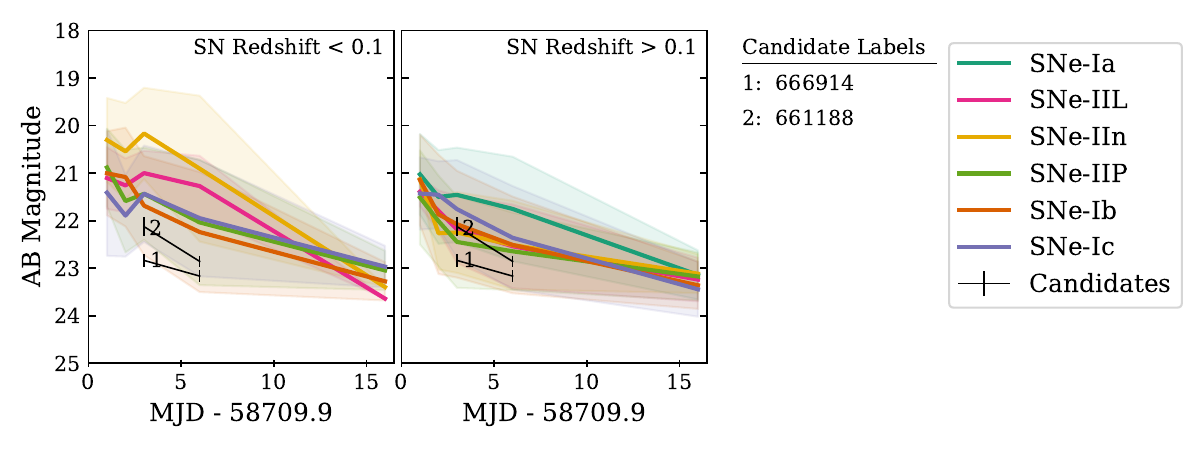}
\includegraphics[width=1.0\linewidth]{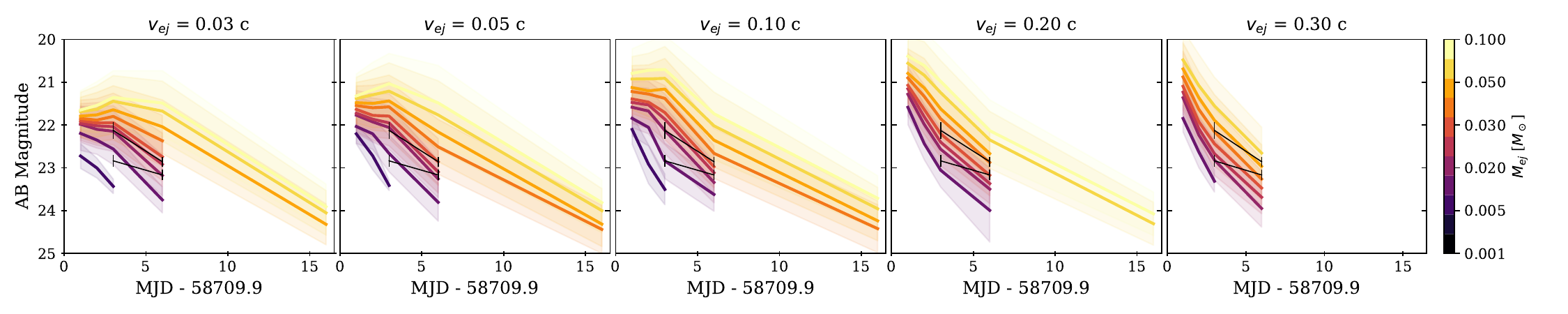}
\caption{Light curves ($i$-band) of objects passing the selection criteria. All light curves are simulated and averaged to determine the mean light curve and $1\sigma$ confidence level contours. For the simulated SNe, $z=0.1$ is used as the cutoff between low and high redshift. Other classes of SNe did not pass the selection criteria with high enough frequency to be accurately represented in the figure.
Simulated KNe light curves averaged over $X_{lan}$ are shown in the bottom panel.
Our candidates passing criteria 1-9 are overlaid for qualitative assessment.
\label{fig:lightcurves}}
\end{figure*}

Here we present the results of applying our real-time selection criteria to {\tt SNANA}-simulated SNe and KNe light curves.
We stress that this approach of representing the expected signal and background samples by applying selection criteria to the light curves provides our best understanding of the characteristics of the objects present in the final candidate sample.
We utilize our simulated light curves to quantify the expected number of remaining SNe in the final candidate sample, to determine the detection efficiencies of all available KN models, to understand the light curves of objects passing our selection criteria, to place upper limits on the physical properties of the merger, and to forecast our discovery potential in future follow-up observations.
In Section \ref{sec:discussion}, we use all these pieces of information to inform a discussion of efficient follow-up strategy and on the dynamics of the merger.

Table \ref{tab:cuts} lists the number of candidates surviving each criterion enforced during our real-time analysis.
We also show that the number of candidates remaining after all selection criteria is consistent with the expected background SNe in these follow-up observations.
The ML classification of our candidates found no potential KNe remaining in our final sample.
Furthermore, because the {\tt PSNID+FRC} classifier performed with a false-positive-rate of 0.01, a remaining candidate would be identified as a KN at the $3\sigma$ confidence level.
This low false-positive-rate of the classifier effectively reduces the SNe background to zero objects, which will prove to be essential for claiming an association between a GW signal and a candidate counterpart in subsequent optical follow-up observations.

A second result of this analysis is the detection efficiency of 329 independent KN models as they would appear in our DECam observations.
The top panel of Figure \ref{fig:KN_efficiency} shows the efficiency of each model after Criterion 1 was placed.
Criterion 1, which requires a single Type-1 detection, can be thought of as assuring the maximum brightness of the objects is greater than the $5\sigma$ limiting magnitude of the observations.
The middle panel displays the detection efficiencies after all criteria up to the ML classification, and the bottom panel of the figure shows the detection efficiencies after the ML classification occurs. 
The blue and red boxes in the panels identify the best-fit model components of the emission from AT2017gfo \citep[]{Drout1570, Kilpatrick1583}, the optical counterpart for GW170817, at distances consistent with GW190814 and accounting for the environmental conditions of our follow-up observations.
We find that low $M_{ej}$ and high $X_{lan}$ yield an optical signature that would be difficult to detect in our DECam observations.
At the same time, we note that our selection criteria limit our ability to detect KNe models with low $v_{ej}$ and high $M_{ej}$.
Physically, the light curves of these models fade more slowly than other KN models and are more similar to some SN models, which leads to class confusion at the ML stage. 

A third product of this analysis is a prediction of the average light curves of the objects that pass our selection criteria.
In Figure \ref{fig:lightcurves}, we overlay the measured $i$-band magnitudes of our candidates on the average light curves of simulated objects passing the same selection criteria.
In the top row, we consider our candidates in the context of SNe.
The high redshift SNe pass our selection criteria because their photometric host-galaxy redshift and uncertainty are consistent with the LVC distance posterior at the $3\sigma$ level.
As shown in Figure \ref{fig:lightcurves}, these high-redshift SNe very closely resemble our candidates in terms of light curve properties: the fading rates of the light curves over the 16 nights and the apparent magnitudes are quite similar.
The bottom panel compares our candidates to KN models.
Each KN light curve is the average across the full range of $X_{lan}$, since this parameter was found to have the smallest effect on light curve shape--it does however affect the color, but we only show monochromatic light curves in the figure.
{This averaging is subject to the non-uniform grid of models and the parameterization of $X_{lan}$ in log space, however, we reiterate that this parameter has the smallest effect on the light curve shapes shown in the figure.}

The KN models as a class fade much more quickly than our candidates, which results in many of them becoming too faint to detect in our observations 16 nights after the merger.
An understanding of the light curves for a potential KN and for the expected background is essential for choosing an efficient observing strategy, which we will discuss in Section \ref{sec:discussion}.
We note again that KN light curves from models with low $v_{ej}$ and high $M_{ej}$ fade the slowest out of all KN models, and at a rate comparable to the faster-fading SN models in the top panel of Figure \ref{fig:lightcurves}.
This observation only applies to optical emission in the $i$ and $z$ filters, and we are unable to speculate on the generalization of this behavior to other wavelength ranges.
This behavior of this subset of KN models poses the greatest confusion to our ML classifier as a result of the light curve similarities.

\begin{figure*}
\centering
\includegraphics[width=1.0\linewidth, trim={0 2cm 0 3cm},clip]{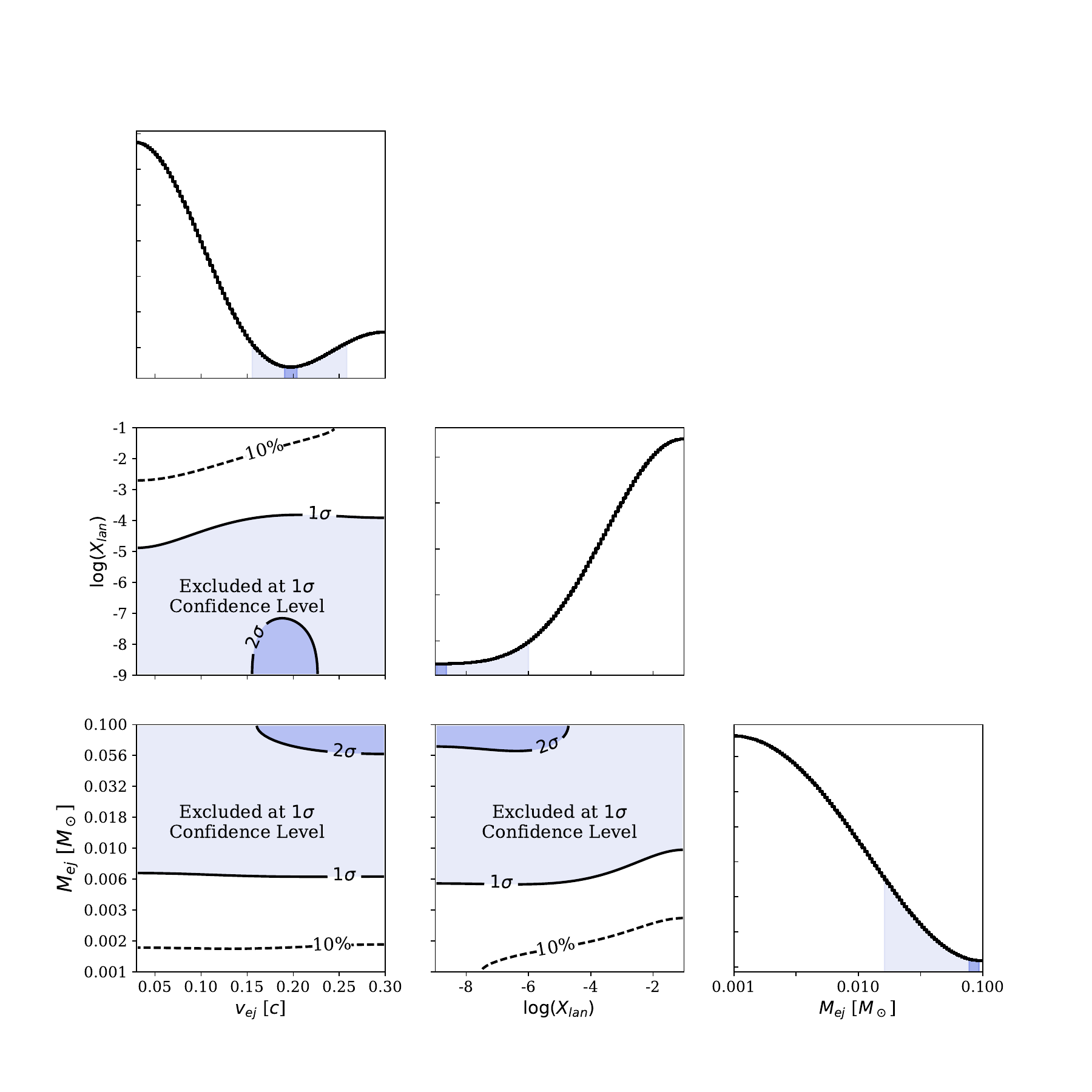}
\caption{Exclusion contours in the ejecta mass ($M_{ej}$), velocity ($v_{ej}$), and lanthanide fraction ($X_{lan}$) parameter space at the $1\sigma$ and $2\sigma$ levels.
The posterior distributions of the three parameters are shown as histograms in the rightmost plot of each row.\label{fig:KN_contours}}
\end{figure*}

Using the fact that a KN was not detected in these observations, we can translate our KN detection efficiencies and expected background rates into upper limits on merger properties.
We estimate the properties of a KN that would go undetected from a Bayesian standpoint:
\begin{align}
P( \textrm{KN}_i | n_{\textrm{cand}}) = \frac{P( n_{\textrm{cand}} | \textrm{KN}_i ) \times  P(\textrm{KN}_i)}{P(n_{\textrm{cand}} )}.\label{eq:bayes}
\end{align}
In Equation \ref{eq:bayes}, KN$_i$ refers to an individual KN model and $n_{\textrm{cand}}$ is the number of candidates detected in the observations. 
In this analysis, $n_{\textrm{cand}} = 0$, though we present the generalized formalism.
The likelihood in Equation \ref{eq:bayes} can be explicitly written as
\begin{equation}
\begin{aligned}[c]
P(n_{\textrm{cand}} | \rm{KN}_i) = \quad & \varepsilon_i \times \textrm{Poisson}(n_\textrm{cand} - 1 | B) \quad + \\ & (1 -  \varepsilon_i) \times \textrm{Poisson}(n_\textrm{cand} | B),
\end{aligned} \label{eq:likelihood}
\end{equation}
where $\varepsilon_i$ represents the detection efficiency of KN$_i$ and $B$ represents the expected SN background, both of which are determined after all selection criteria have been applied.
The Poisson distribution utilized in Equation \ref{eq:likelihood} has an expectation value of $B$~objects and yields the probability of detecting $n_{\textrm{cand}} -1$ or $n_{\textrm{cand}}$~objects.
This formulation is motivated by summing the probability that a KN is detected and the remainder of the candidates are a realization of the predicted SN background with the probability that a KN is not detected and all detected candidates are a realization of the predicted SN background.
In Equation \ref{eq:bayes}, $P(\rm{KN}_i)$ is the prior distribution of KN models, which we make uninformative by assigning equal probability to each model in the non-uniform grid.
The denominator can be evaluated directly by computing
\begin{align}
P(n_{\rm{cand}}) = \sum_{i \in \rm{KN models}} P(n_{\textrm{cand}} | \rm{KN}_i) \times P(\rm{KN}_i),
\end{align}
which can be interpreted as a probability normalization constant. Thus, the posterior distribution of KN models given the non-detection in this analysis can be estimated by setting $n_{\rm{cand}}=0$ and sampling the likelihood space.
The results of this sampling are displayed in Figure \ref{fig:KN_contours}.

From these observations and sensitivity analysis, we report
our constraints on candidate counterpart ejecta properties in Table \ref{tab:constraints}.
To determine the likelihood of a physical KN parameter rather than an individual model in our non-uniform model grid, we perform a three-dimensional linear interpolation between the model efficiencies in the space of $\log(X_{lan})$, $M_{ej}$, and $v_{ej}$.
We note that this linear interpolation is justified by the smoothness of adjacent points in the grid of efficiencies in Figure \ref{fig:KN_efficiency}.
These results are less constraining than what would be obtained using the KN efficiencies and expected backgrounds after Criterion 1, but this is only the case when $n_{\rm{cand}} = 0$.
In this specific case the first term in Equation \ref{eq:likelihood} is zero, which leads to a cancellation of the background in Equation \ref{eq:bayes}, so the effect of the selection criteria only manifests through reducing KN detection efficiencies.
In general, reducing the SN background will produce better constraints.

\begin{deluxetable}{ccc}
\tablehead{\colhead{Ejecta Property} & \colhead{$1\sigma$ Constraint} & \colhead{$2\sigma$ Constraint}}
\startdata
$M_{ej}$ & $< 0.016$~$M_\sun$ & $< 0.07$~$M_\sun$ \\
$v_{ej}$ & $\not\in[0.16c, 0.26c]$ & $\not\in[0.18c, 0.21c]$\\
$X_{lan}$ & $> 10^{-5.92} $ & $> 10^{-8.56} $ \\
\enddata
\caption{Constraints on counterpart ejecta properties of the candidate NSBH merger GW190814. 
These constraints are derived by interpolating the grid of efficiencies in Figure \ref{fig:KN_efficiency} for each of the \citet{Kasen2017} KN models and applying the Bayesian formalism presented in Section \ref{sec:sensitivity_results}.
This calculation utilized an uninformative prior by assigning equal probability to each point in the KN ejecta parameter space.
\label{tab:constraints}}
\end{deluxetable}

It is worth noting here that previous analyses have demonstrated that derived constraints can depend on the models employed in the analysis \citep[]{coughlin19}, and furthermore that the discretized grid of model parameters can affect the constraints as well \citep[]{dietrich2020new}.
For this specific event, optical light would be emitted by tidally-stripped NS material, which motivated our choice of models focused on the ejecta properties.
By not using a model tied to the dynamics of the system, we marginalize over the dependencies on these features of the merger and focus our analysis on directly observable characteristics.
To account for the non-uniform spacing of our grid of model parameters, we performed a three-dimensional linear interpolation of the efficiencies in Figure \ref{fig:KN_efficiency} when performing the Bayesian analysis.
In the Bayesian analysis, we assigned an equal probability to each point in the parameter space of our models.
While not all ejecta parameter combinations may be equally likely given the NSBH-nature of GW190814, we believe our uniformed prior is well-motivated given the mass of the lighter object involved in the merger.
In the event that the object truly was a 2.6~$M_\sun$ NS, we believe all values in the ranges of $M_{ej}$, $v_{ej}$, and $X_{lan}$ are physically accessible under the right dynamical conditions.

To show the benefit of selection criteria that reduce the SN background in GW follow-up observations, we perform simulations of follow-up observations at several points in this analysis.
After each criterion, we take the expected SN background and KN detection efficiency for the blue component of a GW170817-like KN, and calculate the significance level at which that KN would be identified as the counterpart.
Assuming a nearly complete coverage of the GW alert localization area, we report the fraction of follow-up observations where an association at the $1\sigma$, $2\sigma$, $3\sigma$, and $4\sigma$ confidence level would be possible in Figure \ref{fig:gw_forecasting} as functions of the remaining SN background.
Without placing any selection criteria, less than 3\% of DECam follow-up observations can be expected to identify the counterpart at the $3\sigma$ confidence level.
Conversely, with the selection criteria and ML classification developed in this analysis, approximately 95\% of follow-up observations are expected to be able to identify a counterpart in the DECam observations at the $4\sigma$ confidence level.

\section{Discussion}
\label{sec:discussion}

Our optical follow-up observations of the first candidate NSBH merger GW190814, simulations of transients in the localization area, and accompanying sensitivity analysis serve as powerful tools moving forward in the field of multimessenger astronomy.
In this analysis, we presented several key results: the quantification of the expected background, the development of tailored selection criteria, an understanding of KNe efficiency in the observations, an understanding of the light curves of objects in our final candidate sample, upper limits on KN counterpart properties, and the forecasting of our discovery potential using the methods developed here.
In this section, we first compare our results to previous analyses of this merger, and then we utilize our results to inform a discussion of merger dynamics and efficient follow-up strategy.

\subsection{Comparisons to Previous Analyses}
\label{sec:previous_work}

In this subsection, we highlight the differences between our approach and those presented in other analyses and follow-up observations of GW190814.
While G20 analyzed the public DECam observations discussed in this work, multiple teams performed independent observations.
V20 observed GW190814 using MegaCam / CHFT.
The V20 observations utilized the $g$, $i$, and $z$ bands reaching $5\sigma$ limiting magnitudes of $\sim23$~mag on nights 1, 2, 3, 4, 6, 7, 8, and 20 following the merger.
The imaging covered 69\% of the total integrated probability area, as the 1~sq. degree FoV of the imager limited the feasible area to cover each night.
E20 utilized the several observatories and filters to image the 90\% localization area including the Gravitational wave Optical Transient Observer, the Visible and Infrared Survey Telescope for Astronomy, the Very Large Telescope, the Asteroid Terrestrial-impact Last Alert System, and Pan-STARRS1.
They reach limiting magnitudes comparable to DECam on a significant fraction of the localization area and distribute a cadence similar to the DECam and MegaCam cadences across their network of observatories.
M20 performed a galaxy-targeted search within the 50\% localization area on nights 1 and 2 following the merger with the Magellan Baade telescope.
They reach a $3\sigma$ $i$-band limiting magnitude of 22.2 mag.
Lastly, W20 utilized the DDOTI wide-field robotic imager on the first two nights covering the merger.
They cover the full localization area to $\sim18$~mag in the $w = r + 0.23(g-r)$ band.

The characteristics of the different datasets collected, such as imaging depth, observing cadence, sky-area covered, and image quality shaped the analyses performed by the counterpart search teams.
W20 was able to detect transients to $\sim18$~mag, meaning KN-like optical signatures at the distance of GW190814 would be too faint to detect.
For this reason they are unable to place constraints on counterpart properties that are competitive with the groups employing deeper optical imaging.
M20 obtained deep imaging, but only targeted galaxies with the 50\% localization area (70\% of the galaxy-weighted probability).
While they calculate that KN-like counterparts with more than $0.03$~$M_\sun$ would be too faint to detect in their observations, without covering the full 90\% localization area, they cannot place constraints above the 90\% confidence level.
We, G20, V20, and E20 covered high fractions of the 90\% localization area and utilized telescopes and images powerful enough to detect potential counterparts at the distance of GW190814.



No group reports an EM counterpart, and G20, V20, and E20 use their observations to place constraints on the properties of the merger.
G20 fixes the distance of the merger to the mean value of 267~Mpc and finds $M_{ej} > 0.05$~$M_\sun$.
They also consider the viewing angle of the merger in their constraints, which enters in our analysis through the line-of-sight component of the ejecta velocity.
V20 finds slightly tighter constraints on the ejecta mass of a potential EM counterpart (0.015 $M_\sun$), though their analysis fixes $v_{ej}$ to $0.2c$, which we show in this work is disfavored at the $2\sigma$ confidence level.
We suggest this choice of disfavored ejecta velocity is the cause of the comparatively tighter constraints reported by V20.
E20 reports that KN-like counterparts with $M_{ej} > 0.1$~$M_\sun$ are excluded at the 90\% confidence level.
They arrive at this result by utilizing the limiting magnitudes of their observations  and the expected magnitudes of KN models (similar to the work of G20 and V20) at the distances distributed according to the luminosity distance posterior of GW190814 from the LVC (similar to this work).

The characteristic distinguishing the work presented here from the analyses of all other groups is the extent of the sensitivity analysis used to understand KN detection efficiencies in the observations.
G20 and V20 choose a handful of representative fixed distances for the KN and assess whether the the apparent magnitude of a particular model would be brighter than the $5\sigma$ magnitude limit in the band of the observations.
This approach does not consider the effect of the selection criteria applied to the candidates to rule out all objects on the KN model efficiencies, nor does it accurately marginalize over the LVC distance posterior for the merger.
The simulations developed for our work fully incorporate the effects of our real-time selection criteria, the full posterior of luminosity distances, and enable us to place meaningful constraints without fixing any KN parameters.
Understanding the effects of the selection criteria placed during a real-time search on the set of detectable counterpart configurations is essential for accurately constraining the physical properties of the potential optical counterpart.
The approach demonstrated in this work has been utilized in \citet{Doctor_2017}, \citet{me:)}, \citet{de2020zwicky}, \citet{kasliwal2020kilonova}, and \citet{collaboration2020desgw}, and has been facilitated by the development of code bases such as \texttt{simsurvey} \citep[]{simsurvey}.
We advise that future analyses employ this approach in GW counterpart searches and population studies.

\subsection{Merger Dynamics}
\label{sec:discussion_merger}

The optical signature from a NSBH merger is highly dependent on the dynamics of the system and the characteristics of the compact objects involved \citep[]{dynamics1, Bauswein_2013, Radice_2017, Radice_2018}.
For a KN-like signature to be emitted, the NS would need to be tidally disrupted to produce light-emitting ejecta.
Therefore, the spins and masses of the coalescing bodies, which determine the degree of tidal disruption of the NS, are intimately linked to the optical signature \citep[]{Capano2020}.

At the $2\sigma$ confidence level, we were able to exclude counterparts with $M_{ej}$ $>$ 0.07~$M_\sun$.
Thus, only a small fraction of the NS material was ejected.
We also exclude counterparts with $X_{lan}$ $<$ $10^{-8.56}$ at the $2\sigma$ level.
The constraint on this quantity is $10^{-5.92}$ at the $1\sigma$ level, indicating that higher $X_{lan}$ are favored overall, and that in the most probable case, any ejecta produced would have been rich in heavy elements.
This richness could result from the small (if any) amount of NS material ejected in the merger, as the majority of the material would be synthesized into heavy elements by the gravitational potential in close proximity to the BH.
A final result from this sensitivity analysis that can be used to infer properties of the merger is the non-detection of a KN-like counterpart.
Since our KN detection efficiency decreases with $M_{ej}$, the lack of an observation of a KN in this merger event suggests a small or nonexistent amount of ejected material.
The DECam observations are therefore consistent with the NS retaining structural integrity until it passed the radius of the last stable circular orbit.


The physical parameters of the merger most closely tied to the potential tidal disruption of the NS, and hence the optical constraints derived in this analysis, are the mass ratio $q \equiv M_{BH}/M_{NS}$, the magnitude of the final BH spin $\chi$, the radius of the neutron star $r_{ns}$, and the chirp mass $\mathcal{M}$.
The ejected mass increases with decreasing $M_{BH}$, increasing $\chi$, and $r_{ns}$ (harder EOS).
From numerical simulations, the upper limit to disk formation is a mass ratio of  $\sim 3 - 5$ \citep{lattimer_properties_2019, Pannarale_2014, PhysRevD.99.103025, PhysRevD.98.081501}. 
For a fixed BH mass, as the NS mass increases, a larger BH spin is required to produce a massive disk. 
The reason is that higher black hole spin decreases the last stable circular orbit radius, allowing a higher mass NS, generally more compact, to reach its disruption radius and thus leave the disrupted NS matter remaining in orbit.
Holding the NS mass fixed, increasing the BH mass increases the  gravitational radii, and higher spins are needed to bring the last circular orbit radius in below the disruption radius.
Binaries with low mass ratios and high BH spins maximize the chance of massive disk formation. 
Based on our observations, the spins, masses, and alignments of the merging bodies disfavor tidal disruption of the NS.
In their recently release parameter estimation of the merger, the LVC determined $q = 0.11$, $\chi= 0.28$ of 0.28, and $\mathcal{M} = 6.1$~$M_\sun$ \citep{Abbott_2020_0814}.
This spin and mass ratio would lead to small amounts of tidal disruption of the NS, and would be consistent with the lack of an accretion disk, the lack of an accompanying gamma-ray burst, and the lack of a detection a KN-like counterpart.
Therefore, the constraints on NS ejecta properties derived from the DECam observations in this analysis are consistent with the LVC parameter estimation of the merger dynamics.

\subsection{Implications for Follow-up Observation Strategy}
\label{sec:discussion_strategy}

\begin{figure}
\centering
\includegraphics[width=1.0\columnwidth]{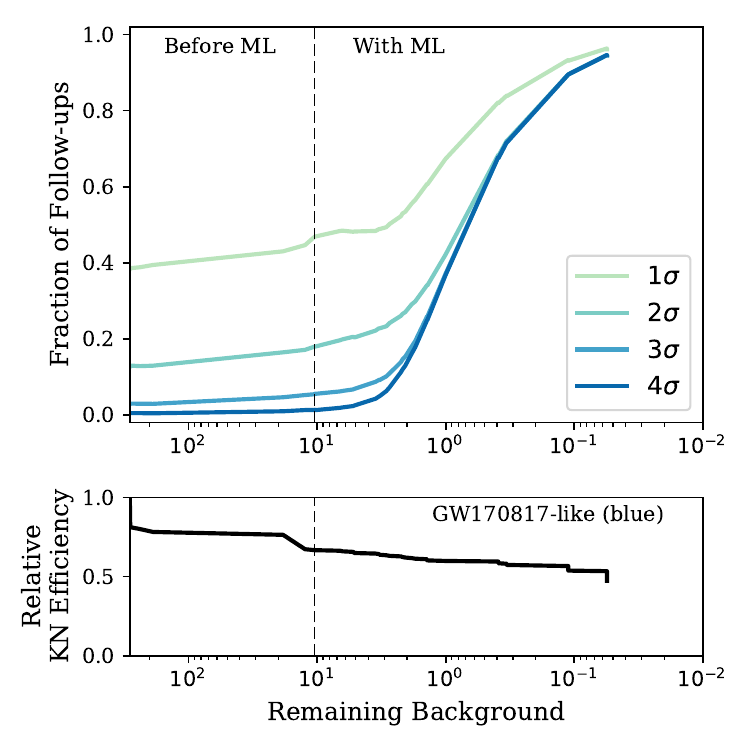}
\caption{The effect of applying selection criteria on follow-up observation sensitivity and KN efficiency. \textit{Top:} The fraction of follow-up observations expected to result in a KN detection of given significance as a function of the remaining background.
\textit{Bottom:} Efficiency for the blue component of a GW170817-like KN at the distance of GW190814 as a function of the remaining background.
\label{fig:gw_forecasting}}
\end{figure}

We find two aspects of our observing strategy for this event to be highly efficient and recommend their use in follow-up observations going forward.
Namely, our observing cadence and exposure time, and specifically how we tailored them to the conditions of the event, were essential in detecting the large number of candidates published by teams using the public DECam data. 

The cadence of our observations was well-suited for the detection of a KN-like optical signal.
By triggering DECam immediately after the LVC alert, and by repeating observations on the next three nights following the merger, we increased our chances of detecting a rapidly changing object.
Furthermore, our choice to include epochs on nights 6 and 16 after the merger enabled the characterization of light curves for longer-lived transient objects in that part of the sky.
This choice proved to be essential in systematically eliminating fading objects unassociated with the GW signal.
While a KN is not expected to be bright at these later epochs, the detection of any potential candidate on these nights can be used as evidence to exclude the object.

In this sensitivity analysis, the availability of light curves spanning a 16 day interval with 6 observing epochs enabled the development of a powerful ML-based photometric classifier.
Our \texttt{PSNID + RFC} approach was able to effectively eradicate the SN background in these observations, and we expect a similar performance in subsequent follow-up observations with similar cadences.
The benefit of devoting resources to background reduction is a key point we seek to make.
Figure \ref{fig:gw_forecasting} demonstrates how reducing the SN background dramatically increases the probability that optical follow-up observations will associate a candidate with the LVC alert at a statistically significant confidence level.
In this figure, we select a single KN model for simplicity and consider its detection efficiency as a function of the remaining SNe background as we apply the selection criteria in this analysis.
The ``Before ML'' section references the real-time selection criteria of this analysis while the ``With ML'' section varies the \texttt{PSNID + RFC} operating threshold to move along the ROC curve of Figure \ref{fig:classification} towards regions of higher purity.
Again, the performance of this classifier and the possibility to reduce the SN background are primarily determined by the cadence chosen by the observing team.
Reducing the SNe background using the techniques of this work leads to the possibility of associating a detected transient with the merger at the $4\sigma$ confidence level $95\%$ of the time in identical follow-up observations.

The exposure time of the DECam images was dynamically varied in response to changing observing conditions.
On the first two nights, when a potential KN-like counterpart would be expected to be near peak brightness, we opted for a shorter exposure time to tile the area twice quickly.
This choice enabled us to rule out moving objects while maintaining enough depth to detect a KN-like object positioned at the estimated distance of GW190814.
In subsequent nights, we increased the exposure time such that we would be sensitive to fainter objects, since a KN-like object would be expected to fade by $\sim0.5$~mag per day \citep{Kasen2017}.
While the choice to vary the exposure time introduces non-uniformity in the image quality of the DECam data, it is useful for maximizing the probability of detecting a rapidly-fading transient on each night of observation.
Furthermore, we note that this variance of image quality over the course of the observations necessitates the use of detailed simulations of the follow-up observations for quantifying constraints.
Our choice of exposure times resulted in the deepest optical observations of the entire 90\% localization region on each night DECam was operated compared to all follow-up teams.
Thus, the observing strategy described in this work is a useful baseline for future DECam follow-up observations.

The chances of detecting a potential counterpart were greatly improved by SOAR spectra being obtained for the most interesting candidates.
While we were able to achieve high accuracy machine-learning-based photometric classification of the objects in our sample in this work, the success of that approach requires the availability of several nights of photometric observations.
In the real-time portion of the observations, the spectroscopic component of the search is essential.
The use of SOAR enabled us to confidently exclude $\sim20$\% of our most promising candidates on the first few nights of the observations.

We see this spectroscopic efficiency as an aspect of gravitational wave counterpart identification that can be dramatically improved given the resources of the astronomical community, for example with the use of wide--field multi--object spectroscopy \citep{2019BAAS...51c.310P}.
In cases where the 90\% localization is larger than what one telescope can cover in a single night, the fraction of sky area covered by the astronomical community is another improvable trait of the follow-up strategy.
As we look forward to the increased sensitivity in Observing Run 4 and consequent increased alert frequency, synergy among follow-up teams will be integral to the association of gravitational waves with their electromagnetic counterparts.
Distributed and coordinated observations among follow-up teams will be essential, and furthermore the sharing of observation metadata to improve sensitivity and forecasting studies will benefit the field as a whole.
For further discussion of these topics, see \citet{Coughlin2020}

\section{Conclusion}
\label{sec:conclusion}


In response to the first high confidence alert of gravitational radiation from a neutron star--black hole merger GW190814, we triggered the 4m Blanco Telescope / Dark Energy Camera and obtained the deepest coverage of the entire 90\% localization area.
Our observations took place on 6 nights over the first 16 nights following the merger, and each night the Dark Energy Survey Gravitational Wave Search and Discovery Team published candidate counterpart objects to the astronomical community \citep[]{GCN_1, GCN_2, GCN_3, GCN_4, GCN_5, GCN_6, GCN_7, GCN_8, GCN_9, GCN_10, GCN_11, GCN_12, GCN_13, GCN_14, GCN_15, GCN_16}.
The entire localization area was within the Dark Energy Survey footprint, enabling the use of six years of previous images and complete host galaxy catalogs in our search for a counterpart.
In an offline analysis following the conclusion of observations, all candidates were excluded based on light curve properties, photometric redshifts of the host galaxies, or a machine-learning classification approach developed specifically for this work.
We present the results of the real-time follow-up observations and accompanying sensitivity analysis here.
Using detailed simulations of supernovae and kilonovae matched to our observing cadence and conditions, we quantify the expected supernova background, develop selection criteria that effective remove that background, and calculate kilonova efficiency resulting from the selection criteria.
The non-detection of an electromagnetic counterpart in our data implies that a potential counterpart had $M_{ej}$ $<$ 0.07 $M_\sun$, $v_{ej} < 0.18c$ or $ v_{ej} > 0.21c$, and $X_{lan} > 10^{-8.56}$ at the $2\sigma$ confidence level.
These analysis components enabled us to also characterize the typical light curves of supernovae and kilonovae that would appear in our observations, set constraints on the properties of an undetected kilonova, and forecast the sensitivity of follow-up observations like this one going forward.
We utilize these results to inform a discussion of the dynamics of the merger and efficient gravitational wave follow-up strategy.

\section*{Acknowledgments}
R. Morgan thanks the LSSTC Data Science Fellowship Program, which is funded by LSSTC, NSF Cybertraining Grant \#1829740, the Brinson Foundation, and the Moore Foundation; his participation in the program has benefited this work. 
F.O.E.\ acknowledges support from the FONDECYT grant nr.\ 1201223.

This material is based upon work supported by the National Science Foundation Graduate Research Fellowship Program under Grant No. 1744555. Any opinions, findings, and conclusions or recommendations expressed in this material are those of the author(s) and do not necessarily reflect the views of the National Science Foundation.

Funding for the DES Projects has been provided by the U.S. Department of Energy, the U.S. National Science Foundation, the Ministry of Science and Education of Spain, 
the Science and Technology Facilities Council of the United Kingdom, the Higher Education Funding Council for England, the National Center for Supercomputing 
Applications at the University of Illinois at Urbana-Champaign, the Kavli Institute of Cosmological Physics at the University of Chicago, 
the Center for Cosmology and Astro-Particle Physics at the Ohio State University,
the Mitchell Institute for Fundamental Physics and Astronomy at Texas A\&M University, Financiadora de Estudos e Projetos, 
Funda{\c c}{\~a}o Carlos Chagas Filho de Amparo {\`a} Pesquisa do Estado do Rio de Janeiro, Conselho Nacional de Desenvolvimento Cient{\'i}fico e Tecnol{\'o}gico and 
the Minist{\'e}rio da Ci{\^e}ncia, Tecnologia e Inova{\c c}{\~a}o, the Deutsche Forschungsgemeinschaft and the Collaborating Institutions in the Dark Energy Survey. 

The Collaborating Institutions are Argonne National Laboratory, the University of California at Santa Cruz, the University of Cambridge, Centro de Investigaciones Energ{\'e}ticas, 
Medioambientales y Tecnol{\'o}gicas-Madrid, the University of Chicago, University College London, the DES-Brazil Consortium, the University of Edinburgh, 
the Eidgen{\"o}ssische Technische Hochschule (ETH) Z{\"u}rich, 
Fermi National Accelerator Laboratory, the University of Illinois at Urbana-Champaign, the Institut de Ci{\`e}ncies de l'Espai (IEEC/CSIC), 
the Institut de F{\'i}sica d'Altes Energies, Lawrence Berkeley National Laboratory, the Ludwig-Maximilians Universit{\"a}t M{\"u}nchen and the associated Excellence Cluster Universe, 
the University of Michigan, NFS's NOIRLab, the University of Nottingham, The Ohio State University, the University of Pennsylvania, the University of Portsmouth, 
SLAC National Accelerator Laboratory, Stanford University, the University of Sussex, Texas A\&M University, and the OzDES Membership Consortium.

The UCSC team is supported in part by NASA grant NNG17PX03C, NSF grant AST-1815935, the Gordon \& Betty Moore Foundation, the Heising-Simons Foundation, and by fellowships from the David and Lucile Packard Foundation to R.J.F.

Based in part on observations at Cerro Tololo Inter-American Observatory at NSF’s NOIRLab (NOIRLab Prop. ID 2012B-0001; PI: J. Frieman), which is managed by the Association of Universities for Research in Astronomy (AURA) under a cooperative agreement with the National Science Foundation.

Based on observations obtained at the Southern Astrophysical Research (SOAR) telescope, which is a joint project of the Ministério da Ciência, Tecnologia, Inovação e Comunicações (MCTIC) da República Federativa do Brasil, the U.S. NSF's NOIRLab, the University of North Carolina at Chapel Hill (UNC), and Michigan State University (MSU).

The DES data management system is supported by the National Science Foundation under Grant Numbers AST-1138766 and AST-1536171.
The DES participants from Spanish institutions are partially supported by MICINN under grants ESP2017-89838, PGC2018-094773, PGC2018-102021, SEV-2016-0588, SEV-2016-0597, and MDM-2015-0509, some of which include ERDF funds from the European Union. IFAE is partially funded by the CERCA program of the Generalitat de Catalunya.
Research leading to these results has received funding from the European Research
Council under the European Union's Seventh Framework Program (FP7/2007-2013) including ERC grant agreements 240672, 291329, and 306478.
We  acknowledge support from the Brazilian Instituto Nacional de Ci\^encia
e Tecnologia (INCT) e-Universe (CNPq grant 465376/2014-2).

This paper has gone through internal review by the DES collaboration.
This manuscript has been authored by Fermi Research Alliance, LLC under Contract No. DE-AC02-07CH11359 with the U.S. Department of Energy, Office of Science, Office of High Energy Physics.

\software{
\texttt{astropy} \citep[]{astropy},
\texttt{matplotlib} \citep{Hunter:2007},
\texttt{numpy} \citep{numpy:2011}, 
\texttt{pandas} \citep[]{pandas:2010},
\texttt{PSNID} \citep[]{psnid},
\texttt{sci-kit learn} \citep[]{scikit-learn},
\texttt{scipy} \citep{scipy:2001},
\texttt{SNANA} \citep{snana}
}

\clearpage

\appendix
\numberwithin{figure}{section}
\numberwithin{table}{section}

\section{Machine Learning Photometric Classification}
\label{app:ml}

The use of machine-learning algorithms in astronomy has grown tremendously over the last decade.
From image classification with convolutional neural networks to anomaly detection in data releases, the practice of making an inference about data based on archival or simulated data is a useful tool for improving analyses.
We take a similar approach in this work by utilizing simulated SNe and KNe light curves to develop a classification scheme for the objects found in the real-time DECam observations of GW190814.

Our training set for the ML classification consisted of simulated light curves matched to the DECam observations of GW190814.
That is, the cadence, the measured zeropoints, noise level in the template images from our observations were all incorporated into the simulated fluxes and uncertainties.
Furthermore, for a light curve to be used in the training set, it also had to pass all criteria leading up to the ML classification step, so all light curves were bright enough to be detectable,  consistent in redshift with the LVC distance posterior, and fainter than 22.5~mag on the final night of the observations.
This training set thus not only matches the characteristics of the data to be classified, but also homogenizes the population such that a classification algorithm is forced to find meaningful characteristics as sources of information.

Of all the simulated light curves passing the selection criteria, 70\% are used in training and 30\% are used for testing. 
As well, the number of SNe and number of KNe light curves in each set are forced to be equivalent in order to prevent representation bias from affecting the classification.
Half of the training set was used for initial hyperparameter tuning of the classifier before being recombined with the remainder of the training set for the final classifier training.

In most cases, time-series data are not straightforward to feed into an ML algorithm.
To give the information a classifier sees structure, a standard practice is to extract features from the time-series data.
An example of this is using the average slope in the $g$-band or the maximum signal to noise in the light curve as representations of the information contained in the full light curve.
Our approach performs feature extraction using a Bayesian light curve fitting tool known as \texttt{PSNID}.
\texttt{PSNID} performs a maximum likelihood fit of several SNe-Ia, SNe-Ibc, and SNe-II to the light curves and outputs goodness of fit metrics as well as the best fit parameters of those templates for each light curve.
This process yields $\sim60$ potential features for informing our classification scheme.
We select a subset of 15 of these features by normalizing each feature to a range of zero to one, calculating the difference between the means of that feature for KNe and SNe, and choosing the 15 features for which this metric is largest.
The chosen features are listed and described in Table \ref{tab:features}.

To determine which of these features are the most important when trying to distinguish KNe versus SNe and how to best use them, we apply a Random Forest Classifier to the \texttt{PSNID} fit outputs.
A random forest is a bagged and boosted ensemble of decision trees.
A decision tree learns how to make splits on features in a dataset by optimizing a pre-specified metric (such as information gain, or entropy).
The constituents of the random forest are made different by having them learn from different subsets of the training data, a process known as ``bagging.''
Furthermore if an individual training light curve is found to significantly hinder (help) the classifications made in the decision trees it is used by, it becomes weighted less (more) in the process of shaping the decision trees.
This weighting scheme is also used on a feature-by-feature basis, such that features that lead to better (worse) performance across the ensemble are weighted more (less) when making classifications.
Both of these are examples of a process called ``boosting.''
In our classifier, we use a forest of 100 decision trees, set the maximum depth of each decision tree to be 5 decisions, and utilize entropy as our metric to be optimized when shaping the decision trees.
The weights applied to the classification features are depicted in the right panel of Figure \ref{fig:classification}.
Finally, once the weighted decision trees are constructed, a test set or real data light curve is passed through each tree, and a probability of KN versus SN is assigned based on the number of trees (and their weights) that vote one way or the other.
Therefore, one can select an operating threshold KN probability above which an object is classified as a KN and below which is classified as a SN.

A minor hiccup exists in the fact that this output probability is not inherently physically meaningful -- it is merely the result of how many trees voted for a KN and how many voted for a SN -- so choosing the best threshold is not immediately obvious.
In our approach we impute physical meaning into the probabilities by calibrating the output KN probability to the false negative rate.
That is, the output probabilities are kept in the same order, but weighted by a monotonic function such that choosing an operating threshold of 0.25 corresponds to having a false negative rate of 0.25.
The false negative rate is equivalent to one minus the true positive rate for a KN.
Therefore, this example chosen threshold means an object classified as a KN will have a 75 \% chance of being a true KN.
After calibrating our probabilities, we choose to operate the classifier at a threshold of 0.198, which leads to a KN purity greater than 0.99, a KN completeness of 0.802, a false positive rate of 0.01, and an accuracy of 0.898.
A confusion matrix showing how frequently a light curve of a given type is classified as a KN, SN-CC, and SN-Ia in the left panel of Figure \ref{fig:classification}.

\begin{figure*}
\centering
\includegraphics[width=0.49\linewidth]{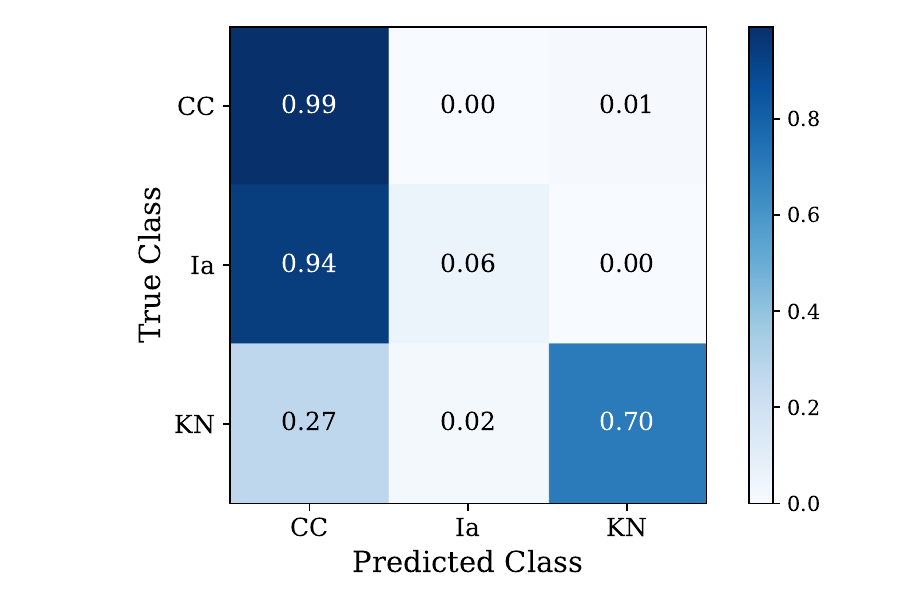}
\includegraphics[width=0.49\linewidth]{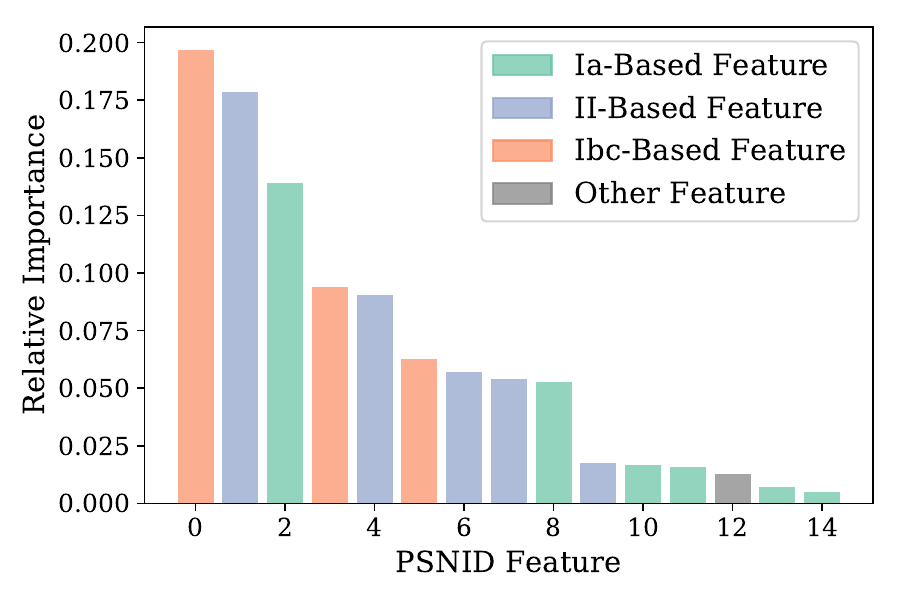}
\caption{Performance of the \texttt{PSNID + RFC} approach. \textit{Left:} A normalized confusion matrix for the different object classes in this classification problem. \textit{Right:} The relative importances of the features used in our approach and from which type of SN template they are derived.
Names and descriptions of each feature are listed in Table \ref{tab:features}.
\label{fig:diagnostics}}
\end{figure*}

\begin{deluxetable*}{cll}
\tablehead{
\colhead{Index} & \colhead{Name} & \colhead{Description} }
\startdata
0 & \tt{FITPROB\_Ibc} & 1 - p-value when fitting a SN-Ibc template to the light curve  \\
1 & \tt{FITPROB\_II} & 1 - p-value when fitting a SN-II template to the light curve \\
2 & \tt{FITPROB\_Ia} & 1 - p-value when fitting a SN-Ia template to the light curve \\
3 & \tt{CHI2\_Ibc} & $\chi^2$ value when fitting a SN-Ibc template to the light curve \\
4 & \tt{CHI2\_II} & $\chi^2$ value when fitting a SN-II template to the light curve \\
5 & \tt{TOBSMIN\_Ibc} & Best-fit MJD of first observability of a SN-Ibc template \\
6 & \tt{TOBSMIN\_II} & Best-fit MJD of first observability of a SN-II template \\
7 & \tt{TOBSMAX\_II} & Best-fit MJD of last observability of a SN-II template \\
8 & \tt{CHI2\_Ia} & $\chi^2$ value when fitting a SN-Ia template to the light curve \\
9 & \tt{LCQ\_II} & Light Curve Quality estimate based on flux error-bar size compared to the SN-II template\\
10 & \tt{LCQ\_Ia} & Light Curve Quality estimate based on flux error-bar size compared to the SN-Ia template\\
11 & \tt{COLORPAR\_Ia} & Best fit color of a SN-Ia template \\
12 & \tt{ITYPE\_BEST} & {\tt PSNID} classification of the light curve \\
13 & \tt{SHAPEPAR\_Ia} & Best-fit shape of an SN-Ia template (describes duration of explosion) \\
14 & \tt{DMU\_Ia} & Deviation in distance modulus from $\Lambda$CDM based on SN-Ia template fit and standardization\\
\enddata
\caption{\texttt{PSNID+RFC} features chosen by comparing the normalized means of KN and SN samples. The index column corresponds to the PSNID Feature number in the right panel of Figure \ref{fig:diagnostics}. \\ 
\label{tab:features}}
\end{deluxetable*}

\bibliographystyle{yahapj_twoauthor_arxiv_amp}
\bibliography{s190814bv}
\end{document}